\documentclass[times,tighten,twocolumn]{aastex631}
\usepackage{multirow}
\usepackage{graphicx}
\usepackage{subfigure}
\usepackage{color}
\usepackage{amsfonts}
\usepackage{mathrsfs}
\usepackage{amssymb}
\usepackage{hyperref}
\usepackage{ulem}

\usepackage{paralist}

\begin{document}

\title{Changing-look Active Galactic Nuclei from the Dark Energy Spectroscopic Instrument. I.Sample from the Early Data }
\author[0000-0001-9457-0589]{Wei-Jian Guo}
\affiliation{Key Laboratory of Optical Astronomy, National Astronomical Observatories, Chinese Academy of Sciences, Beijing 100012, China Email:\href{mailto:guowj@bao.ac.cn,zouhu@nao.cas.cn}{guowj@bao.ac.cn,zouhu@nao.cas.cn}}
\author[0000-0002-6684-3997]{Hu Zou}
\affiliation{Key Laboratory of Optical Astronomy, National Astronomical Observatories, Chinese Academy of Sciences, Beijing 100012, China
Email:\href{mailto:guowj@bao.ac.cn,zouhu@nao.cas.cn}{guowj@bao.ac.cn,zouhu@nao.cas.cn}}

\author[0000-0003-1251-532X]{Victoria A. Fawcett}
\affiliation{School of Mathematics, Statistics and Physics, Newcastle University, Newcastle, UK}

\author[0000-0003-1398-5542]{Rebecca Canning}
\affiliation{Institute of Cosmology $\&$ Gravitation, University of Portsmouth, Dennis Sciama Building, Portsmouth, PO1 3FX, UK}

\author[0000-0002-0000-2394]{Stephanie Juneau}
\affiliation{NSF’s NOIRLab, 950 N. Cherry Avenue, Tucson, AZ 85719, USA}

\author[0000-0002-4213-8783]{Tamara M.~ Davis}
\affiliation{School of Mathematics and Physics, University of Queensland, 4072, Australia}

\author[0000-0002-5896-6313]{David M. Alexander}
\affiliation{Centre for Extragalactic Astronomy, Department of Physics, Durham University, South Road, Durham, DH1 3LE, UK}

\author[0000-0003-4176-6486]{Linhua Jiang}
\affiliation{Kavli Institute for Astronomy and Astrophysics, Peking University, Beijing 100871, People's Republic of China}

\author{Jessica Nicole Aguilar}
\affiliation{Lawrence Berkeley National Laboratory, 1 Cyclotron Road, Berkeley, CA 94720, USA}

\author[0000-0001-6098-7247]{Steven Ahlen}
\affiliation{Physics Dept., Boston University, 590 Commonwealth Avenue, Boston, MA 02215, USA}

\author{David Brooks}
\affiliation{Department of Physics \& Astronomy, University College London, Gower Street, London, WC1E 6BT, UK}

\author{Todd Claybaugh}
\affiliation{Lawrence Berkeley National Laboratory, 1 Cyclotron Road, Berkeley, CA 94720, USA}

\author[0000-0002-1769-1640]{Axel de la Macorra}
\affiliation{Instituto de F\'{\i}sica, Universidad Nacional Aut\'{o}noma de M\'{e}xico,  Cd. de M\'{e}xico  C.P. 04510,  M\'{e}xico}

\author{Peter Doel}
\affiliation{Department of Physics \& Astronomy, University College London, Gower Street, London, WC1E 6BT, UK}

\author[0000-0003-2371-3356]{Kevin Fanning}
\affiliation{The Ohio State University, Columbus, 43210 OH, USA}

\author[0000-0002-2890-3725]{Jaime E. Forero-Romero}
\affiliation{Departamento de F\'isica, Universidad de los Andes, Cra. 1 No. 18A-10, Edificio Ip, CP 111711, Bogot\'a, Colombia}
\affiliation{Observatorio Astron\'omico, Universidad de los Andes, Cra. 1 No. 18A-10, Edificio H, CP 111711 Bogot\'a, Colombia}

\author[0000-0003-3142-233X]{Satya Gontcho A Gontcho}
\affiliation{Lawrence Berkeley National Laboratory, 1 Cyclotron Road, Berkeley, CA 94720, USA}

\author{Klaus Honscheid}
\affiliation{Center for Cosmology and AstroParticle Physics, The Ohio State University, 191 West Woodruff Avenue, Columbus, OH 43210, USA}
\affiliation{Department of Physics, The Ohio State University, 191 West Woodruff Avenue, Columbus, OH 43210, USA}
\affiliation{The Ohio State University, Columbus, 43210 OH, USA}

\author[0000-0003-3510-7134]{Theodore Kisner}
\affiliation{Lawrence Berkeley National Laboratory, 1 Cyclotron Road, Berkeley, CA 94720, USA}

\author[0000-0001-6356-7424]{Anthony Kremin}
\affiliation{Lawrence Berkeley National Laboratory, 1 Cyclotron Road, Berkeley, CA 94720, USA}

\author[0000-0003-1838-8528]{Martin Landriau}
\affiliation{Lawrence Berkeley National Laboratory, 1 Cyclotron Road, Berkeley, CA 94720, USA}

\author[0000-0002-1125-7384]{Aaron Meisner}
\affiliation{NSF's NOIRLab, 950 N. Cherry Ave., Tucson, AZ 85719, USA}

\author{Ramon Miquel}
\affiliation{Instituci\'{o} Catalana de Recerca i Estudis Avan\c{c}ats, Passeig de Llu\'{\i}s Companys, 23, 08010 Barcelona, Spain}
\affiliation{Institut de F\'{i}sica d’Altes Energies (IFAE), The Barcelona Institute of Science and Technology, Campus UAB, 08193 Bellaterra Barcelona, Spain}

\author[0000-0002-2733-4559]{John Moustakas}
\affiliation{Department of Physics and Astronomy, Siena College, 515 Loudon Road, Loudonville, NY 12211}

\author[0000-0001-6590-8122]{Jundan Nie}
\affiliation{Key Laboratory of Optical Astronomy, National Astronomical Observatories, Chinese Academy of Sciences, Beijing 100012, China Email:\href{mailto:guowj@bao.ac.cn,zouhu@nao.cas.cn}{guowj@bao.ac.cn,zouhu@nao.cas.cn}}

\author[0000-0003-0230-6436]{Zhiwei Pan}
\affiliation{Kavli Institute for Astronomy and Astrophysics, Peking University, Beijing 100871, People's Republic of China}
\affiliation{Department of Astronomy, School of Physics, Peking University, Beijing 100871, People's Republic of China}

\author{Claire Poppett}
\affiliation{Lawrence Berkeley National Laboratory, 1 Cyclotron Road, Berkeley, CA 94720, USA}
\affiliation{Space Sciences Laboratory, University of California, Berkeley, 7 Gauss Way, Berkeley, CA  94720, USA}
\affiliation{University of California, Berkeley, 110 Sproul Hall \#5800 Berkeley, CA 94720, USA}

\author[0000-0001-7145-8674]{Francisco Prada}
\affiliation{Instituto de Astrof\'{i}sica de Andaluc\'{i}a (CSIC), Glorieta de la Astronom\'{i}a, s/n, E-18008 Granada, Spain}

\author[0000-0001-5589-7116]{Mehdi Rezaie}
\affiliation{Department of Physics, Kansas State University, 116 Cardwell Hall, Manhattan, KS 66506, USA}

\author{Graziano Rossi}
\affiliation{Department of Physics and Astronomy, Sejong University, Seoul, 143-747, Korea}

\author{Małgorzata Siudek}
\affiliation{Institute of Space Sciences (ICE, CSIC), Campus UAB, Carrerde Can Magrans, s/n, 08193 Barcelona, Spain}

\author[0000-0002-9646-8198]{Eusebio Sanchez}
\affiliation{CIEMAT, Avenida Complutense 40, E-28040 Madrid, Spai}

\author{Michael Schubnell}
\affiliation{Department of Physics, University of Michigan, Ann Arbor, MI 48109, USA}
\affiliation{University of Michigan, Ann Arbor, MI 48109, USA}

\author[0000-0002-6588-3508]{Hee-Jong Seo}
\affiliation{Department of Physics \& Astronomy, Ohio University, Athens, OH 45701, USA}

\author{Jipeng Sui}
\affiliation{Key Laboratory of Optical Astronomy, National Astronomical Observatories, Chinese Academy of Sciences, Beijing 100012, China Email:\href{mailto:guowj@bao.ac.cn,zouhu@nao.cas.cn}{guowj@bao.ac.cn,zouhu@nao.cas.cn}}
\affiliation{School of Astronomy and Space Science, University of Chinese Academy of Sciences, 19A Yuquan Road, Beijing 100049, China}

\author[0000-0003-1704-0781]{Gregory Tarl\'{e}}
\affiliation{University of Michigan, Ann Arbor, MI 48109, USA}

\author[0000-0002-6684-3997]{Zhiming Zhou}
\affiliation{Key Laboratory of Optical Astronomy, National Astronomical Observatories, Chinese Academy of Sciences, Beijing 100012, China Email:\href{mailto:guowj@bao.ac.cn,zouhu@nao.cas.cn}{guowj@bao.ac.cn,zouhu@nao.cas.cn}}




\correspondingauthor{Wei-Jian Guo}
\email{guowj163@gmail.com}


\begin{abstract}
 Changing-look Active Galactic Nuclei (CL AGN) can be generally confirmed by the emergence (turn-on) or disappearance (turn-off) of broad emission lines, associated with a transient timescale (about $100\sim5000$ days) that is much shorter than predicted by  traditional accretion disk models. We carry out a systematic CL AGN search by cross-matching the spectra coming from the Dark Energy Spectroscopic Instrument and the Sloan Digital Sky Survey. Following previous studies, we  identify CL AGN based on $\rm{H}\alpha $, $\rm{H}\beta$, and Mg\,{\sc ii} at $z\leq0.75$ and Mg\,{\sc ii}, C\,{\sc iii}], and C\,{\sc iv} at $z>0.75$. We present 56  CL AGN based on visual inspection and three selection criteria, including 2 $\rm{H}\alpha$, 34 $\rm{H}\beta$, 9 Mg\,{\sc ii}, 18 C\,{\sc iii}], and 1 C\,{\sc iv} CL AGN. Eight cases  show simultaneous appearances/disappearances of two broad emission liness. We also present 44 CL AGN candidates with significant flux variation of broad emission lines but remaining strong broad components.  In the confirmed CL AGN, 10 cases show additional CL candidate features for different lines. In this paper,  we find 1)  a 24:32 ratio of a turn-on to turn-off CL AGN;  2) an upper limit transition timescale ranging from 330 to 5762 days in the rest-frame; 3) the majority of CL AGN follow the bluer-when-brighter trend. Our results greatly increase the current  CL census ($\sim30\%$) and  would be conducive to  explore the underlying physical mechanism.
\end{abstract}

\section{Introduction}
In the unification paradigm, different types of AGN or quasars\footnote{CL quasars refer to the most luminous cases with bolometric luminosities $L_{\rm{bol}}>10^{44}\rm{erg/s}$ but  we do not distinguish based on luminosity in the work and use CL AGN for the full range.}§, are classified by their orientation relative to the line-of-sight, and present significant flux variation (about 10\%-30\%) across the entire electromagnetic spectrum on monthly to annual timescales \citep{Antonucci1993, Vanden2004}. The Broad Line Region (BLR)  is located close to the central supermassive black hole (a few light-days to light-months distance). Ionized gas of the BLR emits Broad Emission Lines (BELs) with a smaller flux variation amplitude than the continuum \citep{Kaspi2000,Du2019}. These BELs (in particular, the Balmer lines) have been found to respond to continuum variations with a time delay. This time delay has been used, in a technique known as reverberation mapping, to study  the geometry and kinematics of the BLR and measure the central black hole mass \citep{Blandford1982,Peterson1993, Bentz2009, Ho2014, Li2018,Zhang2019,Lu2021,Bao2022}. 
The variation of BEL features is therefore an effective tool for studying the formation of the BLR and the evolution of AGN. However, Mg\,{\sc ii} and C\,{\sc iv} are less responsive to the continuum variation than Balmer lines because of their intrinsic properties, such as  weak response, the extended BLR size, or outflows \citep{Richards2011, Sun2018, MacLeod2019, Yu2021}.

The changing-look (CL) phenomenon was originally used to characterize X-ray detected AGN that change between Compton-thick and Compton-thin \citep{Matt2003, Temple2022}.
In the optical band,  CL is a surprising and unusual event in which BELs in AGN spectra appear or disappear (from Type 1 to Type 2 or vice versa) within just a few months or years. In  low-redshift  AGN ($z<0.1$), CL could also refer to the transition between intermediate types (such as Type 1.2, Type 1.5, Type 1.8, and Type 1.9) that is determined by the flux ratio between $\rm{H}\beta$\  and $\rm{H}\alpha$\  or [O\,{\sc iii}] \footnote{ In this paper, we define intermediate type transition as a CL candidate where the broad component is still present.}

Over a hundred CL AGN  have been discovered based on their Balmer line profile transitions \citep{MacLeod2016, Gezari2017, Sheng2017, Frederick2019, Hon2022, Green2022, WangJ2022, Zeltyn2022,chen2023,Yang2023}, and dozens of CL AGN have also been identified based on transitions in Mg\,{\sc ii}, C {\sc iii}], and C\,{\sc iv} \citep{guohengxiao2019, Ross2020,Guo2020}. However, the nature and frequency of CL AGN are still not well understood, and many questions remain unanswered. For instance, the occurrence rate of CL events in AGN and the relationship between CL AGN and normal AGN are unclear. Additionally, the mechanisms responsible for these changes are yet to be fully understood.  Three commonly proposed possible causes are 1) changes in the central gas density due to 
variation of energy radiation intensity from the nucleus, or accelerating outflows  \citep{Shapovalova2010, LaMassa2015, Ricci2022}; 2) rapid increase or decrease of gas density and accretion rate in the compact region originating from tidal disruption events (TDE; \citealt{Blanchard2017, Ruancun2022}; 3) accretion rate changes caused by BLR evolution, accretion disk instability, or temporary accretion events \citep{Esin1997, Elitzur2009, Dexter2011, Elitzur2014, MacLeod2019,sniegowska2019}.  CL AGN may be attributed to various mechanisms or influenced by multiple physical processes.

To better understand the physical mechanism behind the CL phenomenon, it is essential and pressing to conduct further spectral monitoring of previously discovered objects and to additionally search for new CL AGN. There are several effective methods to hunt for CL AGN, including 1) cross-matching AGN spectra in multiple-epoch large-area spectroscopic projects, such as those performed by  \citealt{MacLeod2016, Yang2018, Green2022, WangJ2022}; and 2) identifying potential candidates through follow-up spectroscopic observations, by detecting extremely unusual variability in optical or mid-IR light curves, as seen in studies by  \citealt{Sheng2017, MacLeod2019, Graham2020}. The Dark Energy Spectroscopic Instrument (DESI) offers an excellent opportunity to hunt for CL AGN, as it boasts a large field of view, high spectral resolution, and high data generation efficiency (as described in detail in Section  \ref{sec_data}). Another advantage of DESI is the high sensitivity allowing for the identification of lower-luminosity accretion events.

We aim to explore the physical mechanism behind the CL AGN and improve the completeness of the current sample. To achieve this, we will utilize the spectroscopic data from  DESI and the Sloan Digital Sky Survey (SDSS). The DESI project will provide nearly three million quasar spectra in the next five years, while the SDSS project has already identified  750,414 quasars. By cross-matching the AGN spectra from these two projects, we will be able to compile a large sample of CL AGN and study their behaviors. Previous studies have primarily focused on  Balmer line (H$\alpha$ and H$\beta$) CL AGN at redshifts $z<0.7$ \citep{MacLeod2016, Yang2018, MacLeod2019, Graham2020, Green2022}. Since high ionization BELs are thought to connect the inner region of the accretion disk and the BLR \citep{guohengxiao2019, Ross2020}, we aim to study CL AGN for several major BELs at all redshifts, including H$\alpha$, H$\beta$, Mg\,{\sc ii}, C\,{\sc iii}], and C\,{\sc iv}. Therefore, we will divide the sample into two parts based on redshift ($z \leq 0.75$ and $z > 0.75$) to search for CL AGN in both the rest-frame ultraviolet and optical wavelengths. The results of this study will provide crucial insights into the properties of CL AGN and contribute to our understanding of the AGN population.

The paper is structured as follows. In Section \ref{sec_data}, we describe the  data from the DESI and SDSS projects. Section \ref{sec_sample} outlines the process for selecting our target sample, including the use of both visual inspection and spectral variability definitions to identify CL objects. In Section \ref{sec_results}, we present our findings on the five broad emission line CL behaviors, and upper limit timescales, and discuss potential physical origins. The paper concludes with a summary in Section \ref{sec_summary}. Throughout, we adopt a $\Lambda$CDM cosmology with $H_{0}= \text{67 km s}^{-1} \text{ Mpc}^{-1}$, $\Omega_{\Lambda}= 0.68$, and $\Omega_{m}= 0.32$ as reported by \citealt{Planck2020}.

\section{Data}
\label{sec_data}

\subsection{DESI}
DESI\footnote{\url{https://www.desi.lbl.gov/}}  uses the NOIRLab 4m Mayall telescope (8 $\rm deg^{2}$ field view) at Kitt Peak with the aim to constrain the nature of dark energy and probe cosmological distances through the baryon acoustic oscillation (BAO) technique \citep{DESI_Levi,DESI_Aghamousa,DESI_Aghamousa2,
DESI_Silber,DESI_Miller,DESI_Schlafly,DESI_fba,DESI_expcalc}. DESI is carrying out the largest ever multiobject and high-efficiency spectral survey (5000 spectra in a single exposure) and plans to measure 40 million galaxies and quasars within five years \citep{DESI_Aghamousa, DESI_Zhou2020,DESI2022,DESI_Raichoor,DESI_Allende}. The DESI program will accumulate about three million quasar spectra to measure large-scale structures in the main survey\citep{DESI_Lan, DESI_Hahn,DESI_Ruiz-Macias,DESI_yeche}. Since the limiting magnitude of DESI in the  $r$-band reaches about 23 mag, DESI has the unique opportunity to discover fainter and higher redshift quasars than any previous survey, quadrupling the number of quasars discovered by SDSS \citep{Zou2018, DESI_Chaussidon, DESI_Alexander}. Besides the strengths in quantity and depth, three optical channels (blue: $3600–5900 $\AA, green: $5660–7220$\AA, and red: $7470–9800 $\AA )  also provide an excellent spectral resolution (blue: R $\sim$ 2100, green: R $\sim$ 3200, and red: R $\sim$ 4100; \citealt{DESI_Abareshi}). 

\cite{DESI_Guy} describe the spectroscopic data processing pipeline in detail and references therein present the target selection and validation \citep{DESI_Alexander, DESI_Lan, DESI_Raichoor, DESI_Zhou,Myers2023,DESI_Cooper}. Before the start of the main survey,  data from the Survey Validation (SV) was visually inspected to check the spectroscopic quality and redshift reliability \citep{DESI_EDR,DESI_SV}, improve the standard DESI spectroscopic pipelines, such as the ``Redrock" and the ``afterburner'' codes, and reduce the number of misclassified quasars \citep{DESI_Alexander}. 

With a large sample of quasars and high-quality spectra,  DESI offers unprecedented advantages for identifying  CL AGN. Since DESI has the advantage of observing fainter AGN than SDSS,  more turn-off CL AGN are expected to be discovered. In this project, we used 347,201  quasar spectra and 2,955,168 galaxy spectra reduced by the DESI pipeline (based  on the SPECTYPE from ``Redrock") with $\rm ZWARN =0$ \citep{DESI_redrock_qso,DESI_redrock2023}, indicating reliable spectroscopic redshift measurements, which contains SV (named as ``fuji") and the first 2 months of Year 1 data (named as ``guadalupe", \citealt{DESI_SV}). The data of ``fuji" is published as part of the Early Data Release and  ``guadalupe"  will be published at the same time as Data Release 1 (DR1; \citealt{DESI_Alexander, DESI_Lan, DESI_Raichoor,DESI_EDR}).

\begin{deluxetable}{ccccc}
\renewcommand\arraystretch{1.1}
\tabletypesize{\footnotesize}
\tablewidth{0.35\textwidth}
\tablecolumns{6}
\tabletypesize{\footnotesize}
\tabcaption{\centering  The selection of CL AGN and candidates ($\rm{H}\alpha $, $\rm{H}\beta$, and Mg\,{\sc ii}) from  DESI for $z \leq 0.75$
\label{tab1}}
\colnumbers
\tablehead{
\colhead{Group I} &
\colhead{Group II } &
\colhead{Group III} &
\colhead{$\rm N_{sum}$} &
\colhead{Note} 
}
\startdata
32,038  & 32,038 & 2,955,168  & 3,019,244 & Quasar or galaxies  in DESI  \\
4,239   & 1,814  &  2,547     & 8,600     & Cross-match  with DR16 in 1\arcsec\\
4,073   & 1,742  & 2,181      & 7,996     & Available spectra though SAS \\
\hline
1       & 1      & 0          & 2        &  $\rm H \alpha$ CL AGN\\
21      & 3      & 10         & 34        &  $\rm H \beta$  CL AGN \\
2       & 1      & 3          & 6         &  Mg {\sc ii}  CL AGN \\
24      & 5      & 13         & 42        &  Total CL AGN \\
0.59\%      & 0.28\%      & 0.59\%         & 0.53\%        & CL AGN ratio\\
\hline
5       & 1      &  1         & 7         &  $\rm H \alpha$ CL  candidates \\
13      & 0      &  3         & 16        &  $\rm H \beta$  CL candidates \\
5      & 1       &  4        & 10       &  Mg {\sc ii} CL  candidates \\
23      & 2      &  8        & 33        & Total  CL  candidates\\
0.56\%  & 0.11\% &  0.37\%    & 0.41\%     & CL candidates ratio\\
\enddata
\tablecomments{ Columns: (1) AGN number in group I (DESI quasar \&  DR16 quasar), (2) AGN number in group II (DESI galaxy \&  DR16 quasar), (3) AGN number in group III (DESI quasar  \&  DR16 galaxy), (4) the total number of AGN from group I, II, and III.}
\end{deluxetable}

\begin{deluxetable}{ccccc}
\renewcommand\arraystretch{1.2}
\tabletypesize{\footnotesize}
\tablewidth{0.35\textwidth}
\tablecolumns{6}
\tabletypesize{\footnotesize}
\tabcaption{\centering  The selection of CL AGN and candidates (Mg\,{\sc ii}, C\,{\sc iii}], and C\,{\sc iv}) from  DESI for $z > 0.75$
\label{tab2}}
\colnumbers
\tablehead{
\colhead{Group I} &
\colhead{Group II } &
\colhead{Group III} &
\colhead{$\rm N_{sum}$} &
\colhead{Note} 
}
\startdata
315,163 & 315,163 & 2,955,168 & 3,585,494  & Quasar or galaxies  in DESI \\
66,813  &  8,925  &  2,781    & 78,519   & Cross-match  with DR16  in 1\arcsec \\
64,276  & 8,213   & 2,168     & 74,657   & Available spectra though SAS\\
\hline
2     & 0       &  1        & 3       &  Mg {\sc ii} CL AGN\\
16     & 2       & 0         & 18       &  C {\sc iii}] CL AGN \\
1     & 0       & 0         & 1        &  C {\sc iv}  CL AGN \\
19     & 2      &  1         & 22      &  Total CL AGN \\
0.03\%  & 0.02\% &  0.04\%    & 0.03\%     & CL AGN ratio\\
\hline
4      & 1       &  1        & 6      &  Mg {\sc ii} CL  candidates \\
10      & 0       &  0        & 10       &  C {\sc iii}]  CL candidates \\
5      & 0       &  0        & 5       &  C {\sc iv} CL  candidates \\
19      & 1       &  1        & 21      & Total  CL  candidates\\
0.03\%  & 0.01\% &  0.04\%    & 0.03\%     & CL candidates ratio\\
\enddata
\tablecomments{The columns are the same as Table \ref{tab1}.}
\end{deluxetable}

\vspace{-8mm}

\subsection{SDSS/BOSS/eBOSS}
In addition to the DESI data, we  used the SDSS\footnote{\url{https://www.sdss.org/}} spectroscopic data, which covers a large region of the sky and has millions of spectra of galaxies and quasars \citep{SDSSS_gunn}. The SDSS spectroscopic database provides additional high-quality data for the comparison and validation of the DESI redshift measurement.
The SDSS is carried out on a 2.5m Sloan telescope at Apache Point Observatory and has provided a large database of quasar and galaxy spectra through four-stage projects (SDSS-I, SDSS-II, SDSS-III/BOSS, and SDSS-IV/eBOSS; \citealt{Abazajian2009, Alam2017, SDSS_lyke}). The spectra have a wavelength coverage of 3900-9100 \AA \ or 3600-10400 \AA \ and a resolution of R $\sim$ 2000 for quasars and galaxies \citep{SDSS_Eisenstein, SDSS_Smee}. All the SDSS spectra are reduced with SDSS I/II and BOSS data pipelines \citep{SDSS_Stoughton, SDSS_Bolton}. In this study, we have used 750,414 quasar spectra from the SDSS Data Release 16 Quasar (DR16Q) catalog released by \cite{SDSS_lyke} and 4,930,400 galaxy spectra from the DR16 \citep{SDSS_ahumada}

\subsection{Image survery and Light Curve}

Motivated by the target selection for DESI, the DESI Legacy Surveys\footnote{\url{http://legacysurvey.org/}} team utilized a deep and large area image survey \citep{Dey2019}. The survey comprised three projects: the Dark Energy Camera Legacy Survey, the Beijing-Arizona Sky Survey, and the Mayall z-band Legacy Survey, which  covered approximately 14,000 square degrees of extragalactic sky, including 9,900 deg$^{2}$\ in the NGC and 4,400 deg$^{2}$\ in the SGC \citep{Flaugher2015,DESI_Zhou,DESI_dr9}. The survey was conducted using three optical/infrared bands, reaching approximate AB magnitudes of $g=24.0$, $r=23.4$, and $z=22.5$ \citep{Dey2019}. We used the DESI Legacy Survey photometric magnitude to describe the distribution of the final CL AGN sample.

To test the flux calibration of the target using the long-term light curve of photometry, we utilized various surveys and facilities, including the Catalina Real-time Transient Survey (CRTS; \citealt{Drake2009}), Pan-STARRS1 (PS1; \citealt{Chambers2016}), the Palomar Transient Factory (PTF; \citealt{Law2009}), and the Zwicky Transient Facility (ZTF; \citealt{Masci2019}). First, we obtained the magnitude of pseudophotometry by convolving the target's spectrum with the corresponding filter response function. This process allows us to calculate the observed magnitude in a specific bandpass. By analyzing the consistency between the pseudophotometry magnitude and the observed light curve, we can identify any discrepancies or inconsistencies that may indicate calibration issues or instrument problems affecting the identification of CL AGN, such as SDSS fiber drop. We remove 121 spurious CL AGN to ensure the reliability of the final CL AGN sample in Section \ref{sec_oiii}.

\begin{figure*}[t!]
\centering
\includegraphics[width=0.9\textwidth]{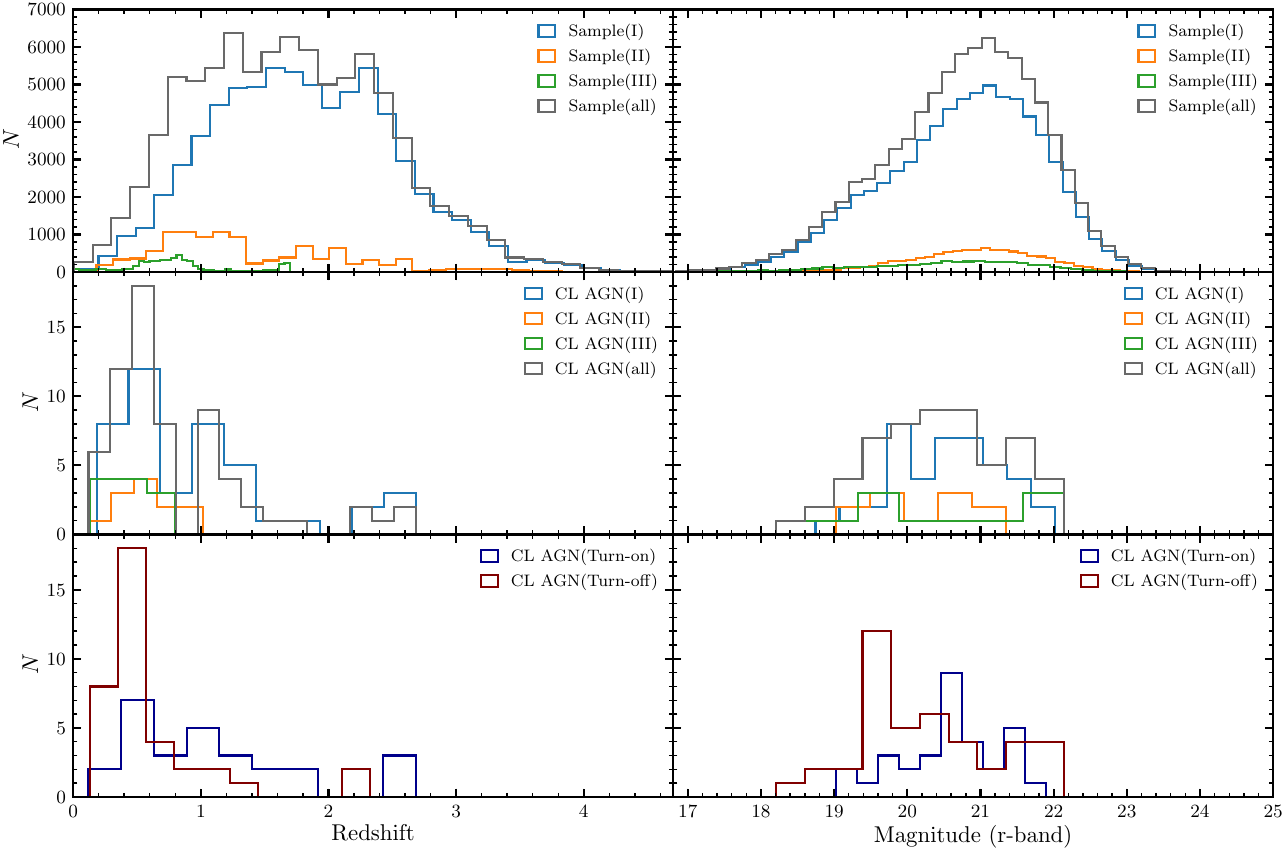}
\caption{The histograms for the redshift (left) and $r$-band magnitude (right) from the DESI Legacy survey are shown. The top panels display the total parent sample (grey) and three cross-matched groups (blue, orange, green). The middle panels represent the distributions of CL AGN found in the three cross-matched groups. The bottom panels show the distributions of CL AGN with the transition state of turn-on (dark blue) and turn-off (maroon). Note that the $r$-band  magnitude refers to the mean magnitude obtained from the image of the DESI Legacy Survey, which encompasses the period from 2014 to 2018.}
\label{fig_redshift}
\end{figure*}

\subsection{Data Preprocessing}
To facilitate the subsequent analyses, including spectrum inspection, defining the change of flux threshold, and sample selection, we corrected the spectra for galactic extinction by using the galactic extinction curve of \cite{Fitzpatrick1999PASP} by assuming $R_{V}=3.1$. After that we shift the spectrum to the rest frame.

Since the spectra of  SDSS and DESI have different wavelength coverages and resolutions,
the flux variation can not be obtained directly by subtraction with two original spectra.
Therefore, we rebin both spectra into the same wavelength grid for flux and its variance (2 \AA\ per pixel) to be able to reliably compare and subtract the SDSS and DESI spectra.

\section{Sample Selection}
\label{sec_sample}

Although DESI has more than one spectrum for some targets, we only use one epoch of DESI for each target and remove any duplicates. This approach does not affect the results of our study as the time interval between the different epochs of DESI spectra is within one year, which is usually shorter than the time scale of most CL objects (for more details, see Section \ref{sec_timescale}).

The sample selection process for CL AGN in this study involves four steps: 1) build AGN parent sample that are both observed by DESI and SDSS; 2) first stage of visual inspection spectra (both SDSS and DESI) to select CL candidates; 3) set the selection criteria  with variability definition to automatically identify of CL AGN in the parent sample; 4) [O\,{\sc iii}]-based calibration for those AGN at $z \leq 0.75$ to remove spurious object; 5) second stage of visual inspection of spectra to remove spurious CL AGN
by comparing the long-term light curve with pseudophotometry obtained.

We define the  parent sample in Section \ref{sec_parent}, describe  the target selection and spectral classification diagnosis by VI in Section \ref{sec_VI}, identify the final CL catalog by the spectral variability definition in Section \ref{sec_definition},
and carry out an [O\,{\sc iii}]-based calibration and pseudophotometry check to remove spurious CL candidates in Section \ref{sec_oiii}.

\begin{deluxetable*}{cccccccc}
\vspace {-8.0mm}
\tablecolumns{8}
\tabletypesize{\footnotesize}
\tabcaption{\centering  The information of five broad emission lines in AGN
\label{tab_lineinfo}}
\colnumbers
\tablehead{
\colhead{$\rm   Line $ } &
\colhead{$\rm  \lambda (\AA )$ } &
\colhead{$\rm Window (\AA ) $ } &
\colhead{$\rm Window\ B1(\AA )$ } &
\colhead{$\rm Window\ B2(\AA )$} &
\colhead{$\rm CL\ AGN$} &
\colhead{$\rm CL\ candidate$} &
\colhead{$\rm Mechanism $} 
}
\startdata
C\  {\sc iv}  &  1548.187, 1550.772 & 1510-1590 & 1460-1510 & 1590-1640 & 1 &  5 & Collisional excitation\\
C {\sc iii}] &  1906.683, 1908.734 & 1850-1970 & 1820-1850 & 1970-2000 & 18 & 10 & Collisional excitation\\
Mg {\sc ii} &  2795.528, 2802.705 & 2750-2850 & 2680-2720 & 2880-2920 & 9 & 16 & Collisional excitation \\
$\rm{H}\beta$    &  4861.333      & 4780-4940 & 4730-4770 & 5030-5080 & 34 & 17 & Recombination\\
$\rm{H}\alpha$    &  6562.819     & 6450-6700 & 6420-6450 & 6750-6790 & 2 & 7 & Recombination\\
\enddata
\tablecomments{Columns: (1) BEL name, (2) BEL wavelength, (3) BEL integration windows,  (4) left continuum windows, (5) right continuum windows, (6) number of the CL AGN founded in DESI, (7) number of the CL candidates founded in DESI, (8)  physical mechanism of the BEL.
}
\end{deluxetable*}

\begin{figure*}
\centering 
\includegraphics[width=0.9\textwidth]{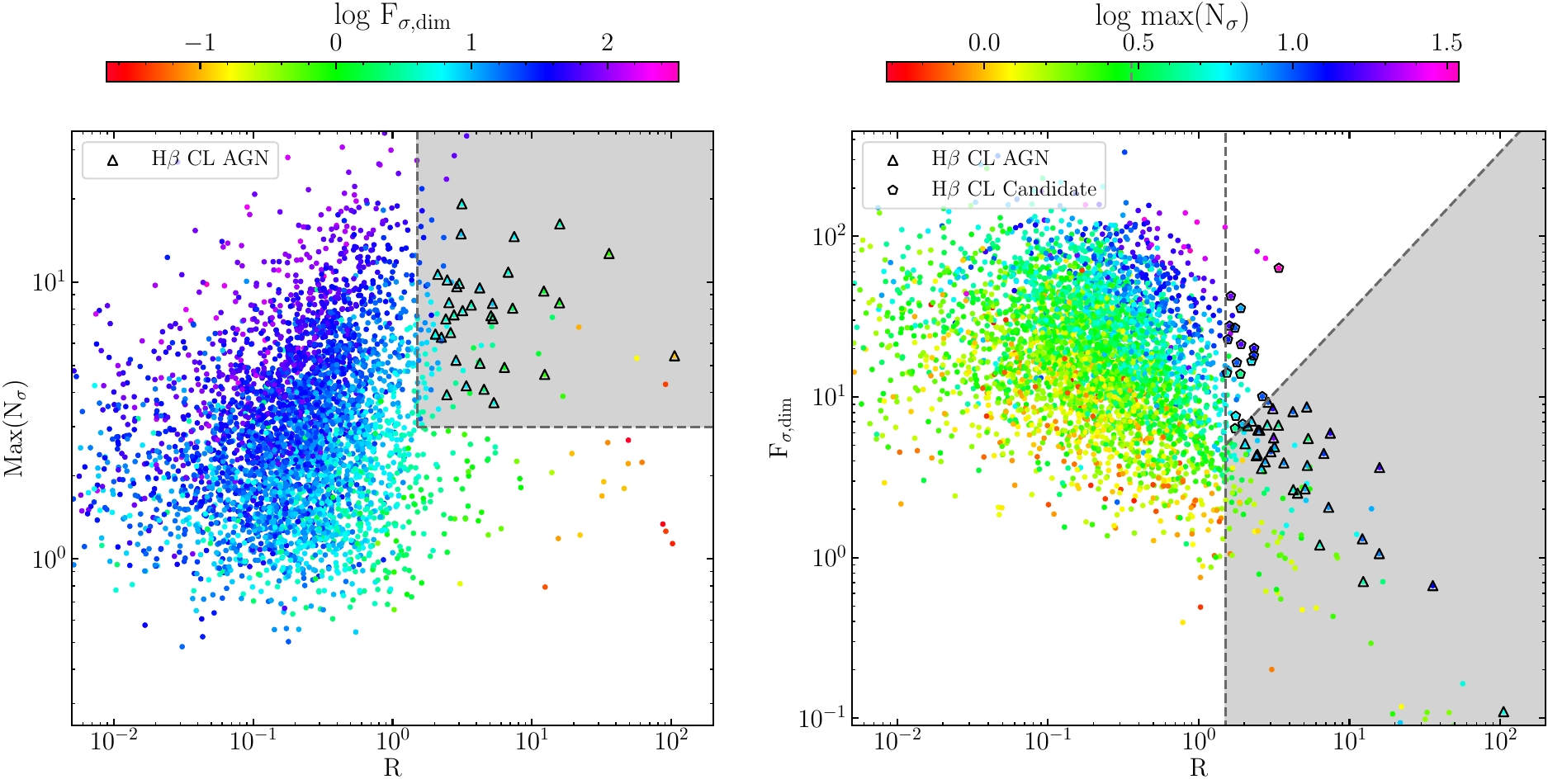}\vspace*{0.4cm}
\includegraphics[width=0.9\textwidth]{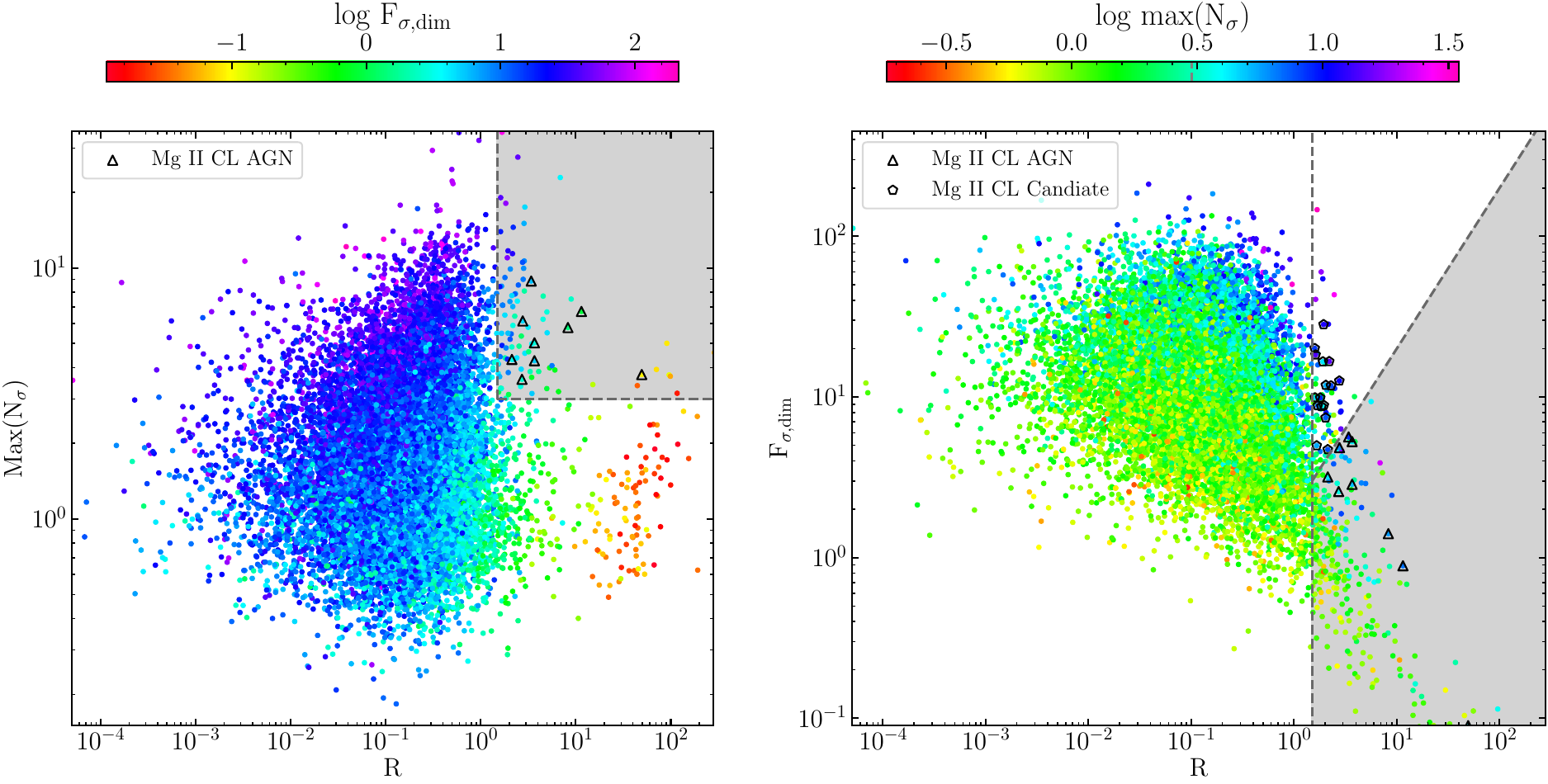}
\caption{ Distribution of CL AGN in  parameter spaces of Max($N_{\sigma}$) vs.  $R$ (left panel) and parameter spaces of  $R$ vs. $F_{\rm \sigma, dim }$ (right panel). The solid scatters represent the measurement  of all AGN parameters ($\rm{H}\beta$ or Mg\,{\sc ii} lies in the the spectrum). The hollow triangles and pentagons represent the final CL AGN  and CL Candidates after the second stage of visual inspection. Gray shading is the parameter thresholds of CL AGN. The dashed gray lines are the corresponding boundary.}
\label{fig_parameter}
\end{figure*}
\vspace{-4mm}

\subsection{Parent Sample}
\label{sec_parent}
We used all the spectra  from the DESI catalog  and SDSS DR16  catalog to systematically search the CL AGN or candidates. To obtain a parent sample of AGN with two epochs of spectra,  we cross-matched the DESI spectra catalog with the SDSS catalog using a 1\arcsec separation, including both quasars and galaxies.  After that, we obtained SDSS spectra of the matched sample through the Science Archive Server (SAS)\footnote{\url{https://dr16.sdss.org/optical/spectrum/search}}. We note here that about 4\% of spectra are missing through SAS downloading, which is possibly due to the data quality. 

Although the spectral classification pipelines of SDSS and DESI perform very efficiently, those objects with indistinct BEL characteristics or low-quality spectra could be leading to a misclassification of the spectral type \citep{DESI_Alexander, DESI_Lan}, especially for CL AGN. As pointed out by many previous works \citep{MacLeod2016, Yang2018, Green2022}, the CL AGN in a low state might be classified as galaxies because the BELs and blue continuum are no longer the dominant features. Therefore, we also included cross-matched results between AGN with galaxies. Specifically, we divided the sample into three groups:  group I (DESI quasar \&  DR16 quasar), group II (DESI galaxy \&  DR16 quasar), and group  III (DESI quasar  \&  DR16 galaxy). Finally, we built up a parent sample containing 82,653  matched AGN spectrum pairs (at least one source in the cross-match is marked as AGN). Figure \ref{fig_redshift} displays the redshift and $r\mbox{-}$band  magnitude  distributions of our parent sample and three different group sample.

Since previous works focus on the  $\rm {H}\alpha $ and $ \rm{H}\beta$ CL phenomenon \citep{MacLeod2016, MacLeod2019, Green2022}, we divided the DESI spectra into two sub-samples according to the redshift to search for: 1) CL AGN defined by $\rm{H}\alpha$, $\rm{H}\beta$, and Mg\,{\sc ii} at $z \leq 0.75$; 2) CL AGN defined by  Mg\,{\sc ii}, C\,{\sc iii}] and C\,{\sc iv} at  $z > 0.75$. Another reason why we set  $z=0.75$ as the boundary between the two sub-samples is that we can use [O\,{\sc iii}] which can be observed in SDSS spectra out to  $z \sim 0.8$  to test the variability definition and estimate the final CL sample purity at  $z > 0.75$, which required the inclusion of both $\rm{H}\beta$ and [O\,{\sc iii}]  in the spectra (for more details, see Section \ref{sec_oiii}). The number of AGN in each of the three groups at $z \leq 0.75$ and  $z > 0.75$  are given in Table \ref{tab1}  and  Table \ref{tab2}, respectively. One might expect that a CL event would always have one epoch classified as an AGN and another as a galaxy (groups II and III).  However, we find that most often both are classified as AGN (group I) since some CL AGN in the dim state may not completely lose the broad components or these objects are  still in the  period of transition. More importantly, the quasar sample were classified by the standard pipeline ``Redrock" or ``QuasarNET" for both SDSS and DESI with about $86\%$\ completeness, which means some AGN, such as host-galaxy dominated AGN, may be contained in galaxy sample \citep{SDSS_lyke,DESI_Alexander}.

\begin{figure*}[!tp]
\centering 
\includegraphics[width=0.49\textwidth]{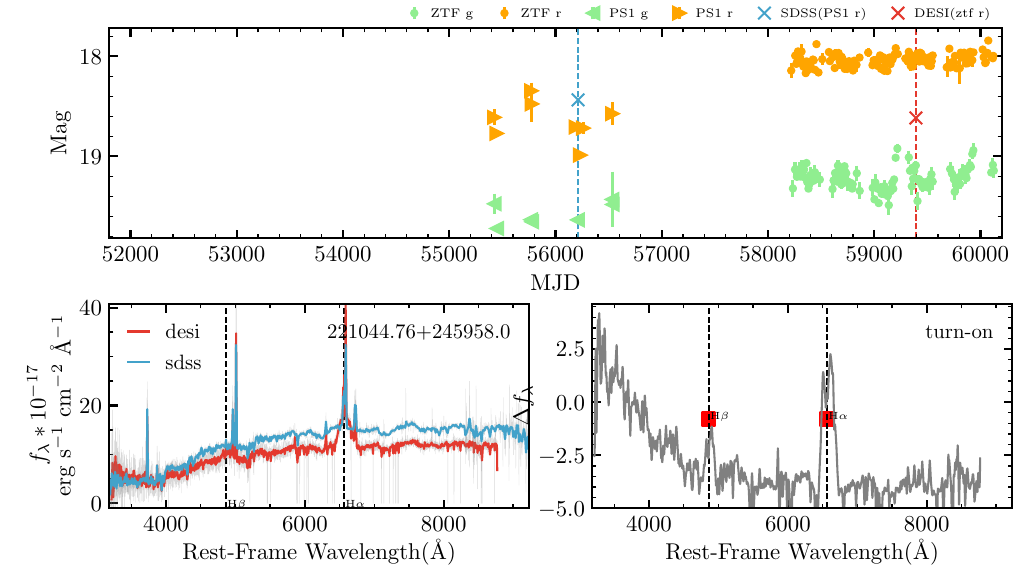}\hspace*{0.01cm}
\includegraphics[width=0.49\textwidth]{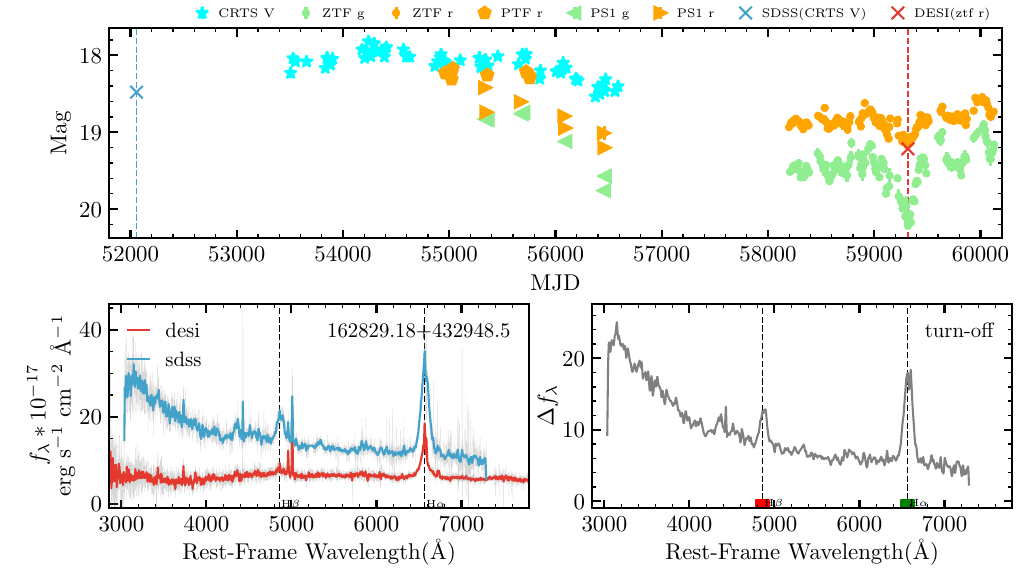}\vspace*{0.01cm}
\includegraphics[width=0.49\textwidth]{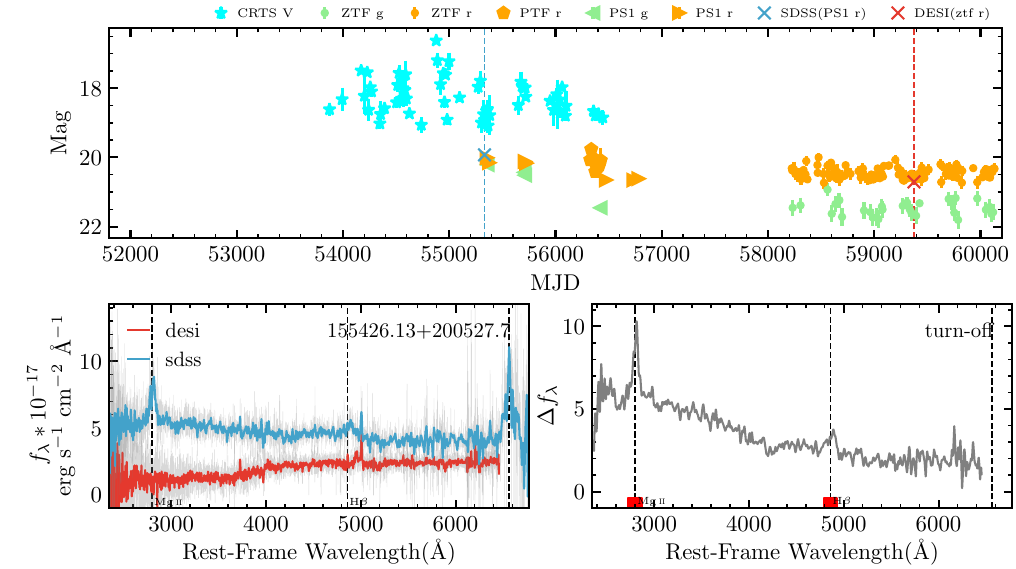}\hspace*{0.01cm}
\includegraphics[width=0.49\textwidth]{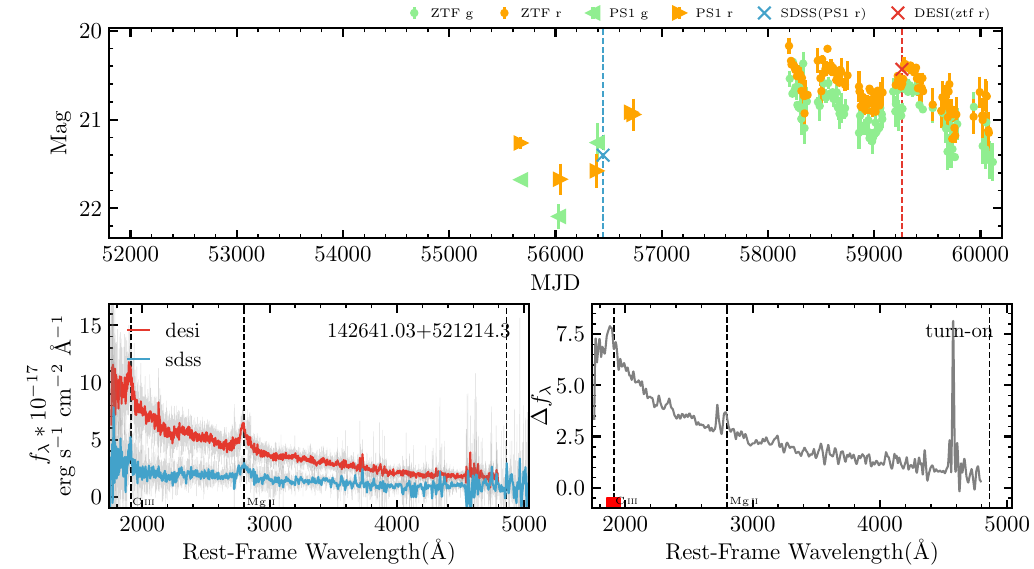}\vspace*{0.01cm}
\includegraphics[width=0.49\textwidth]{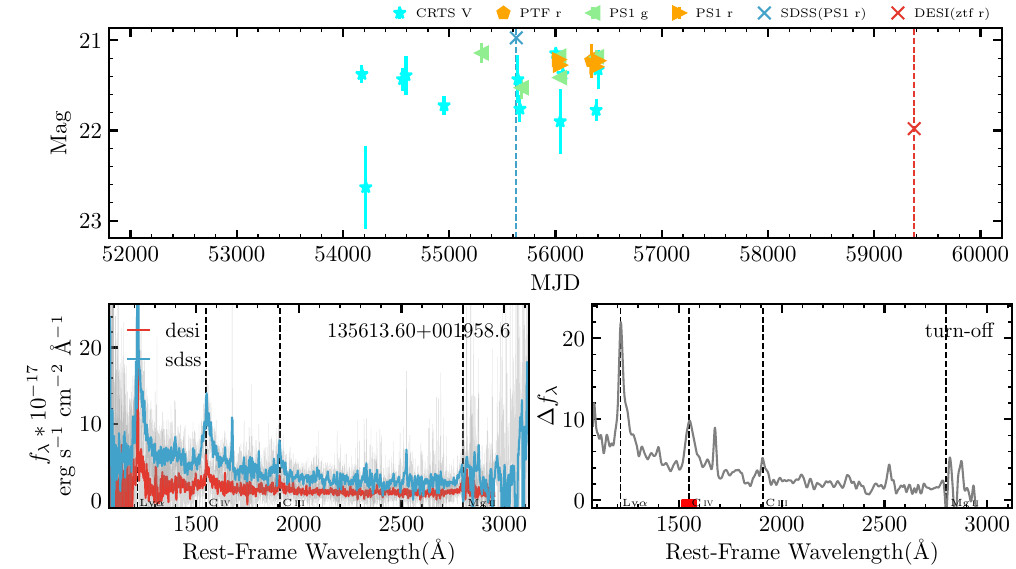}\hspace*{0.01cm}
\caption{Five CL AGN example spectra of $\rm{H}\alpha$ (top left), $\rm{H}\beta$ (top right),  Mg\,{\sc ii} (middle left),  C\,{\sc iii}] (middle right) and  C\,{\sc iv} (bottom) found in DESI. In each figure, the top panels display the spectral pseudophotometry with red (blue) ``x" markers for DESI (SDSS), and the light curves from CRTS (star), PS1 (triangle), PTF (pentagon), and ZTF (dot) when available. The bottom-left panel shows the smoothed SDSS and DESI  represented by blue and red lines respectively (the shaded region corresponds to the original spectrum). The spectra are derived from survey pipelines and  have been corrected for galactic extinction and shifted to the rest frame (without [O\,{\sc iii}]-based calibration). Note that the DESI spectrum covers less flux from the host galaxy than SDSS in 221044.76+245958.0 (Figure \ref{fig_image}.) The grey line in the lower-right panel represents the difference between the two spectra, obtained by subtracting the dim spectrum from the bright one. The black dashed vertical lines indicate the presence of $\rm{H}\alpha$, $\rm{H}\beta$, Mg,{\sc ii}, C,{\sc iii}], C,{\sc iv}, and $\rm Ly \alpha$ lines. The red and green ticks represent the CL line feature and candidate feature, respectively.}
\label{fig_example}
\end{figure*}

\vspace{-2mm}

\subsection{Visual Inspection}
\label{sec_VI}
Previous studies effectively diagnose  CL AGN based on VI of large area sky surveys \citep{MacLeod2016, Yang2018, MacLeod2019, Green2022, WangJ2022}.
Since the appearance or disappearance of BEL may be within a continuous process of spectral enhancement or weakening,  it is a challenge to quantitatively describe the CL behavior with only two randomly sampled spectra.  As mentioned by \cite{Green2022}, a standard CL AGN target selection process starts with a VI to discard those spectra that have poor quality, large measurement errors, or wrong redshift identifications.

In this project, we also carry out  VI of the parent sample for three purposes. Firstly, we need to remove those fake CL AGN whose behavior might be caused by spectrum defects and/or disagreement between the SDSS and DESI redshifts. For example, the absorption associated with the C\,{\sc iv} line (for example, Broad Absorption Line AGN \citep{Rogerson2018}) would impact the accuracy of the total BEL flux. Secondly, since the wavelength region of $\rm{H}\alpha$  contains strong narrow emission lines (narrow $\rm{H}\alpha$ and [N {\sc ii}]), the variation of BEL flux determined by the integration method could be affecting the selection. We inspect all the  $\rm{H}\alpha$  CL candidates to confirm they are indeed CL AGN  in the selection program (see detail in Section \ref{sec_results}). 

Thirdly, the final spectral variability definition that we apply is the automatic checker, which was established based on the results from the VI and run on all the parent samples (see Section \ref{sec_definition}).  In conclusion, the VI check is a crucial step in searching for CL AGN and candidates in large sky surveys, as it helps to remove false positive results and establish a reliable spectral variability definition.

\begin{figure*}
\centering
\includegraphics[width=0.32\textwidth]{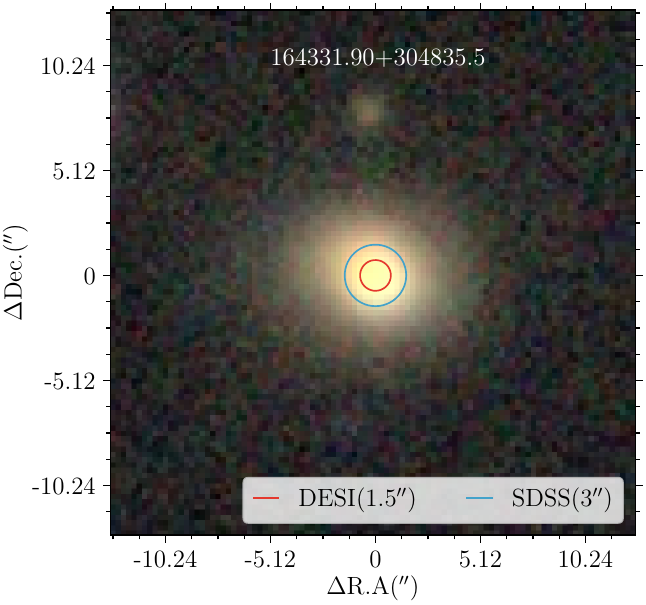}
\includegraphics[width=0.32\textwidth]{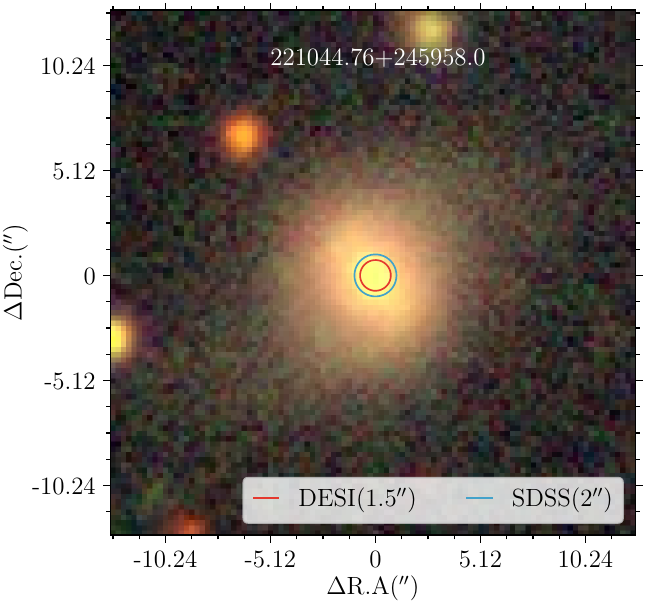}
\includegraphics[width=0.32\textwidth]{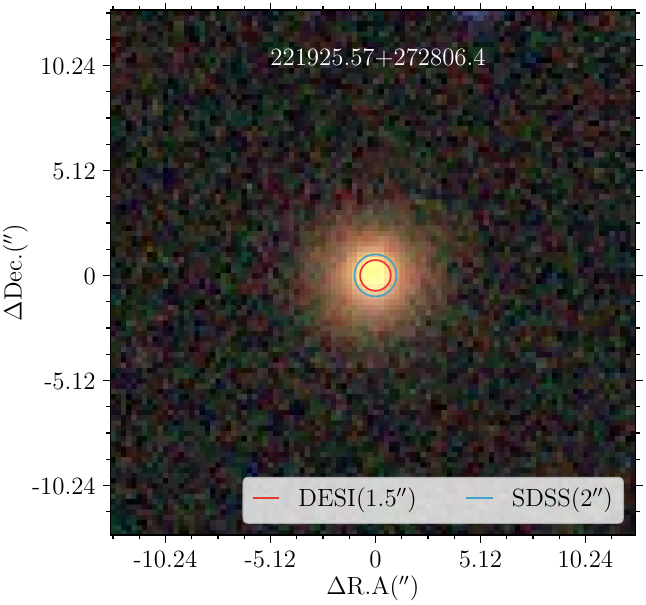}
\caption{The DESI Legacy Image for J164331.90+304835.5 (left), J221044.76+245958.0 (middle), and J221925.57+272806.4 (right). The blue and red circle represent the  fiber diameter of the SDSS and DESI.}
\label{fig_image}
\end{figure*}

\begin{figure*}
\centering 
\includegraphics[width=0.49\textwidth]{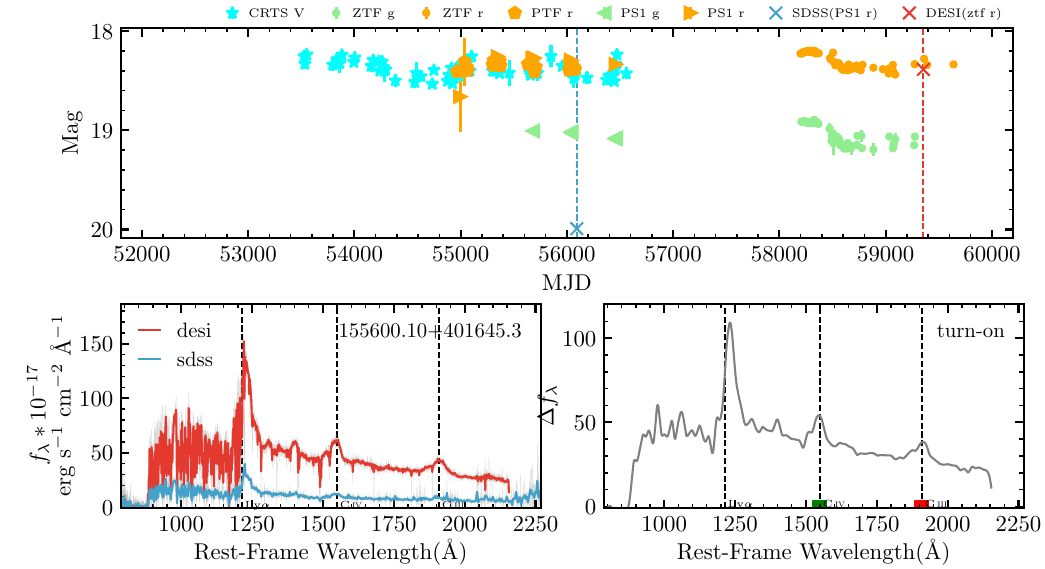}\hspace*{0.01cm}
\includegraphics[width=0.49\textwidth]{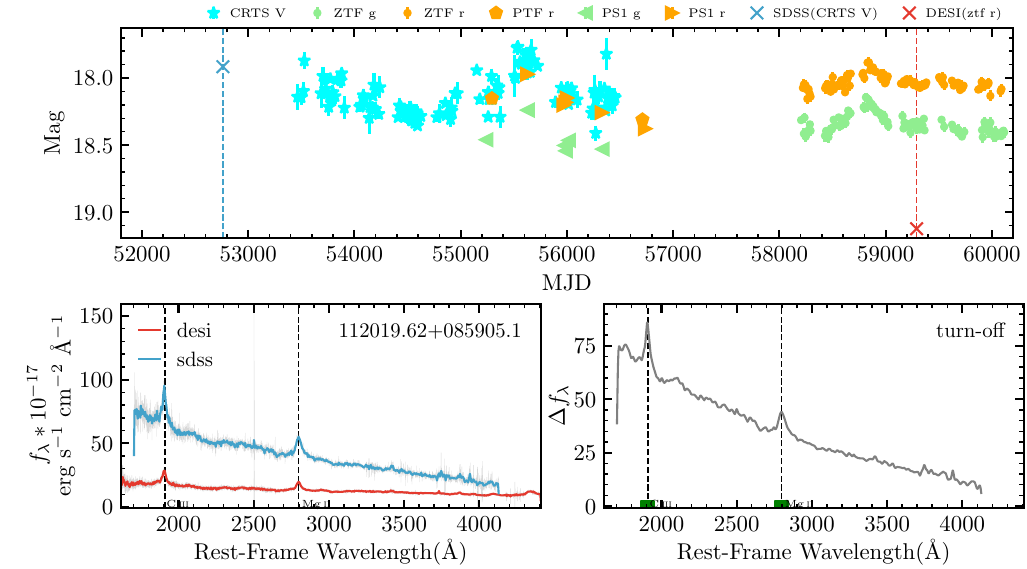}
\caption{Two example spurious CL phenomena due to the SDSS (left) and DESI (right) flux calibration problem. The legends are same as Figure \ref{fig_example}. In the left (right) figure, the SDSS (DESI) spectrum does not match the calculated PS-1 (ZTF) photometric magnitude (``x" markers) in the light curve, while the DESI (SDSS) spectrum aligns with the light curve. This suggests a discrepancy in flux calibration for this spectrum.}
\label{fig_problem}
\end{figure*}

\subsection{Spectral Variability Definition (Selection Criteria) }
\label{sec_definition}

Before defining the variability of the BEL, we require that the median $\rm S/N $ per pixel   is at least greater than one for both the DESI  and SDSS spectrum. In previous studies (e.g. \citealt {MacLeod2016, Yang2018, Graham2020, Green2022, Temple2022}),
a variety of criteria were chosen to describe the CL behavior of the BELs. Based on these previous studies, we adopt three spectrum parameters ($N_{\sigma}$, $R$, and $F_{\rm \sigma, dim }$) to accurately describe the CL phenomenon.

$N_{\sigma}$ is defined as the significance in the variation of the BEL maximum flux and is a widely used definition applied for $\rm H\beta$ CL AGN \citep{MacLeod2019, Green2022}, which is the flux deviation between the dim spectrum and bright spectrum:
\begin{eqnarray}
\label{eq_N}
N_{\sigma}=(f_{\rm bright }-f_{\rm dim } )/ \sqrt{\sigma^{2}_{\rm bright}+\sigma^{2}_{\rm dim} },
\end{eqnarray}
where $f$  and $\sigma$  are the spectral flux and variance respectively in  $\rm erg \ cm^{-2}\  s^{-1} \AA ^{-1}$. \cite{Green2022} smoothed the  $N_{\sigma}(\rm H\beta)$ array, subtracted the $N_{\sigma}(4750\rm \AA)$ by the flux deviation array, and found the maximum relative value of $N^{'}_{\sigma}(\text{H}\beta)=N_{\sigma}(\text{H}\beta)-N_{\sigma}(4750\rm \AA)$. The AGN with $N^{'}_{\sigma}(\rm H\beta)>3$\ are considered as a CL object (see \citealt{Green2022} Section 3.5 for details). Although this determination is effective at determining the ``BEL disappeared or appeared" behavior, they also noticed that their selection criteria were very sensitive to minor BEL changes if the spectrum $\rm S/N $ is sufficiently high, which can be seen in several of the CL objects discovered in \cite{MacLeod2019} and \cite{Green2022}.  

In this paper, we aim to define a variability definition that can recreate our VI results and is also less biased by spectrum $\rm S/N$. We use the maximum value of the smoothed $N_{\sigma} > 3$  objects without the subtraction of the $N_{\sigma}(4750\AA)$  as the first determination of BEL variation to limit the overall variation of both continuum and BELs.

The second parameter ($R$) is defined as the ratio of the integrated BEL flux in the high-state spectrum to the integrated BEL flux in the low-state spectrum, which is used to constrain the total flux change of BELs. As mentioned and inspired by \cite{MacLeod2019}, we further adopt the $R$ value to constrain the overall BEL flux change, which is  defined as:
\begin{eqnarray}
\label{eq_R}
R =(F_{\rm bright }-F_{\rm dim } )/F_{\rm dim },
\end{eqnarray}
where  $F_{\rm bright}$ or $F_{\rm dim}$ is the BEL total flux. Integration and spectral fitting are two widely used tools to obtain the BEL flux in the field of reverberation mapping \citep{Peterson1993,hu2020}. The integration method used in this work provides a quick and easy way to measure the broad emission line flux, which is suitable for the massive AGN dataset released by DESI. We note that the accuracy of  integration may be affected by the host contribution or optical (or UV) FeII lines. By subtracting the continuum as a straight line defined by two continuum windows, the BEL flux is measured by a simple integration method (see detail in Figure 2  of \cite{hu2020}). 
The BEL integration windows are carefully adjusted to ensure the best match with the first stage of VI results. The information on the integration windows is listed in Table \ref{tab_lineinfo}.

If assume that [O\,{\sc iii}] flux is unchanged,  the $R$ value we use is the same as the criteria defined in \cite{Winkler1992}:
\begin{eqnarray}
R =\frac{F_{\rm bright, H\beta }/F_{\rm bright,[O\,{\sc III}]}}{F_{\rm dim, H\beta }/F_{\rm dim,[O\,{\sc III}]}}-1.
\end{eqnarray}
An AGN would be classified as different sub-types if the $\rm{H}\beta$ and [O\,{\sc iii}]  flux ratio change is larger than 1.5, such as from Type 1.2 to Type 1.5 or from Type 1.5 to Type 1.8 (see definition in \cite{Winkler1992} for detail), which we would refer to a CL candidate in our sample. Based on $\rm{H}\beta$, we adopt $R >1.5$ as the CL candidates selection criterion for all BELs. 

The last parameter  $F_{\rm \sigma, dim}$ represents the significance of the weak BEL, which is calculated from the dim spectrum\footnote{For a turn-on CL AGN, SDSS spectrum is at the dim state. If a CL AGN is marked as a turn-off, the dim spectrum will be DESI}:
\begin{eqnarray}
 F_{\rm \sigma, dim} = \sum f_{\rm dim }/ \sqrt{\sum \sigma^{2}_{\rm dim}}.
\end{eqnarray}
Since the narrow component emission lines (such as narrow H$\beta$, [N {\sc ii}]) could exist in the dim spectra to increase  $F_{\rm \sigma, dim}$ value,  we take a different $F_{\rm \sigma, dim}$ value for Balmer lines and other lines.  Another reason for the lower $F_{\rm \sigma, dim}$ value  for other lines is mainly due to a lower S/N for high redshift AGN.

When the S/N is high enough, the AGN with a tiny broad component may be categorized into CL candidates. Compared with the weak component, we care more about the flux variation ratio.  
Thus, in the case of $R >1.5$, we restrict $F_{\rm \sigma, dim}<5$ for  Balmer lines ($F_{\rm \sigma, dim}<3$ for other lines) as the criterion for CL AGN.
As $R$ increases, we lower the limit of  $F_{\rm \sigma, dim}$ and use a straight line to divide CL AGN and candidates with $F_{\rm \sigma, dim }= 3.33R$ for  Balmer lines or $F_{\rm \sigma, dim }= 2.0R$ for other lines, which is plotted in Figure \ref{fig_parameter}.

\begin{figure*}
\centering 
\includegraphics[width=0.49\textwidth]{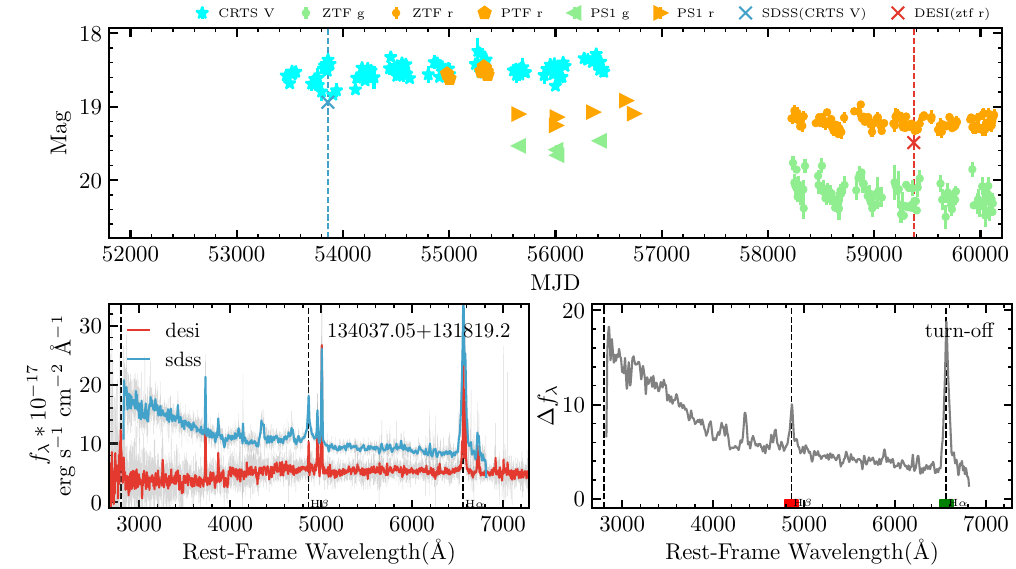}\hspace*{0.01cm}
\includegraphics[width=0.49\textwidth]{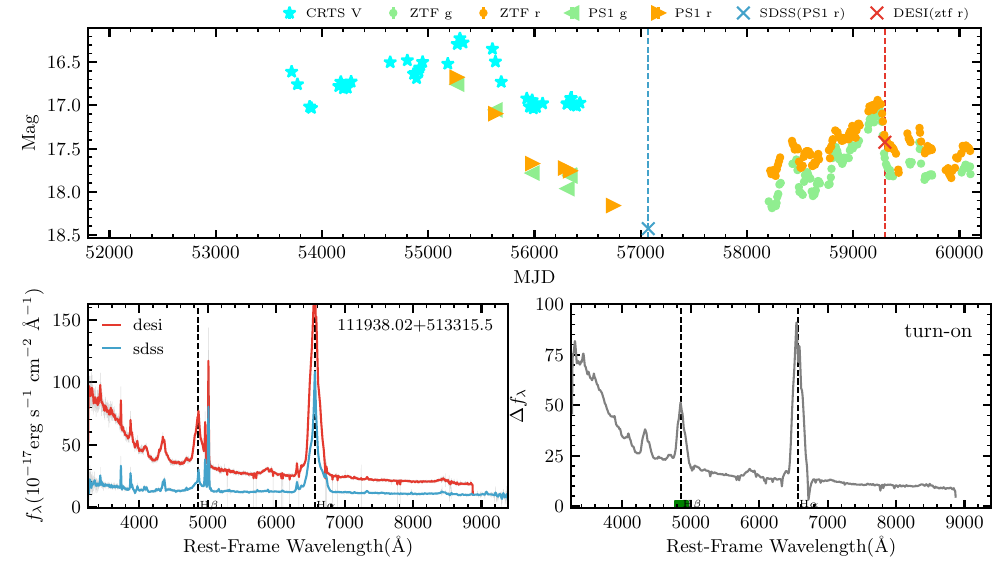}\vspace*{0.01cm}
\includegraphics[width=0.49\textwidth]{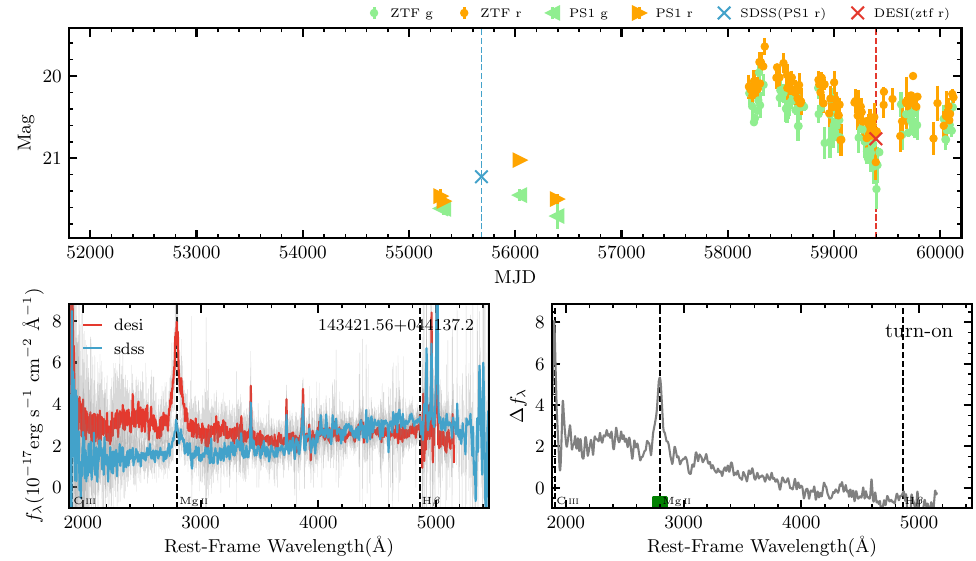}\hspace*{0.01cm}
\includegraphics[width=0.49\textwidth]{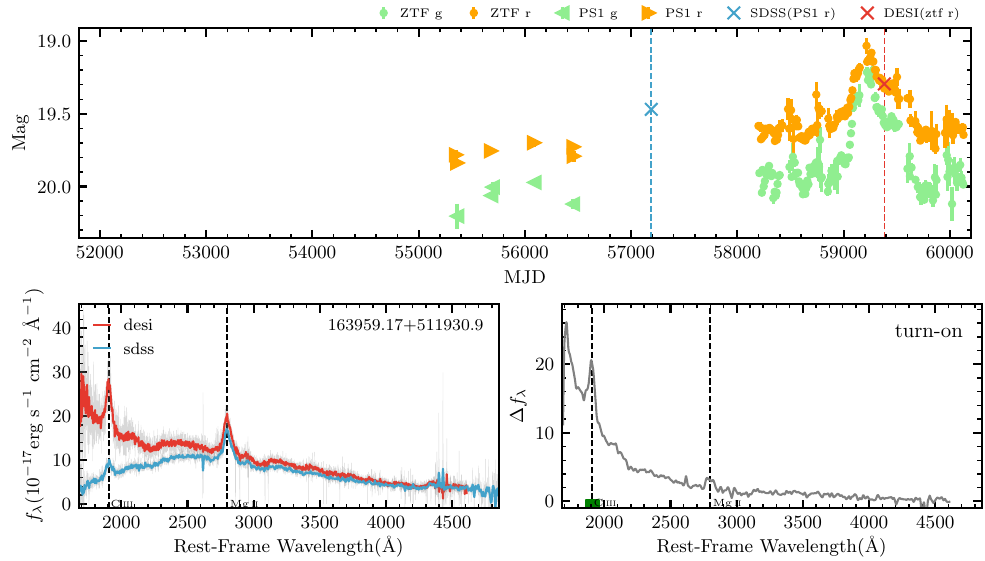}\vspace*{0.01cm}
\includegraphics[width=0.49\textwidth]{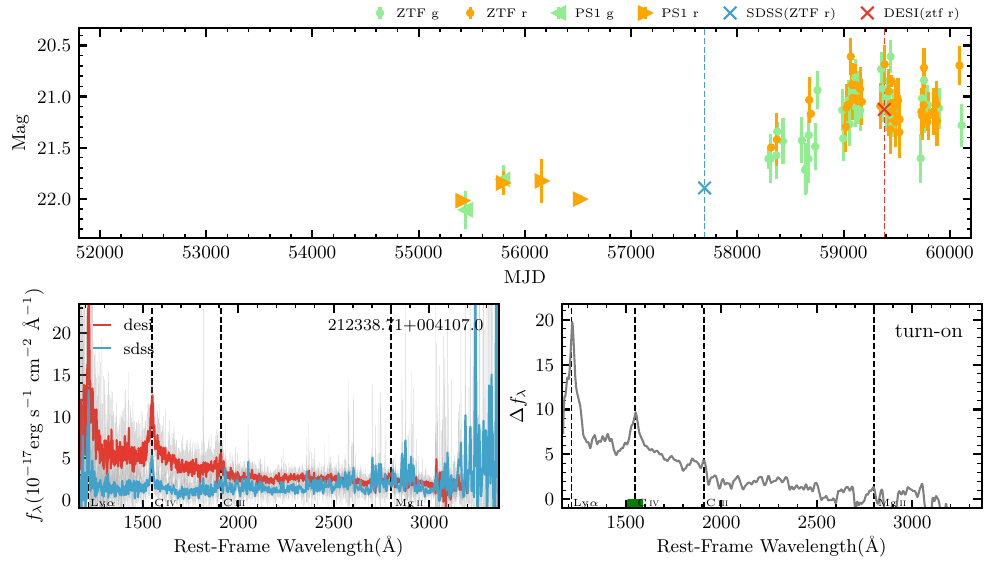}\hspace*{0.01cm}
\caption{Same with Figure \ref{fig_example}, but  showing the CL candidates for
$\rm{H}\alpha, \rm{H}\beta$,  Mg\,{\sc ii},  C\,{\sc iii}], and  C\,{\sc iv} respectively. The complete figures for all  CL candidates are available in their entirety in the online form.}
\label{fig_example4}
\end{figure*}

By combining these three parameters, we can accurately describe the CL behavior of the BELs. As mentioned before, our CL AGN are mainly for the transition  between type 1.x to type  1.9, while CL candidates are mainly an intermediate transition, such as similar to from type 1.2 (1.5) to 1.5 (1.8) for $ \rm{H}\beta$. The final variability definitions of these parameter thresholds are also established based on the results from the VI check and the comparison with previous studies.  The final selection criteria for CL  candidates and AGN are:
\begin{itemize}
    \item  CL candidates:  $\rm Max(N_{\sigma}) > 3$,  $ R >1.5$;
    \item  CL AGN for  $\rm{H}\alpha$ and $ \rm{H}\beta$: $\rm Max(N_{\sigma})  > 3$, \par
     \qquad  $R >1.5$, $F_{\rm \sigma, dim }< 3.33R$; \par
    \item  CL AGN for  Mg\,{\sc ii}, C\,{\sc iii}] and C\,{\sc iv}:  $\rm Max(N_{\sigma})  > 3$,\par 
      \qquad $R >1.5$,  $F_{\rm \sigma, dim }< 2.0R$.
\end{itemize}
It is  important to keep in mind that these parameter thresholds are tentative and may need to be refined or adjusted based on future large datasets. This selection method is expected to be less biased by the spectral $\rm S/N$ and provide a robust and efficient way to identify CL AGN in the DESI project. We also note that spectral decomposition may result in identifying weak broad components even though the  BEL disappeared in VI results. Therefore, regardless of whether the BEL is disappearing or not, CL candidates and  CL AGN may be incorrectly identified due to the coincidence of spectral sampling or weak  BEL flux covered by the host galaxy (this will be discussed in Section \ref{sec_origin}).     

In Figure \ref{fig_parameter} you can see there is a clear correlation between these three parameters, revealed by plotting two parameters and colour-coding the third. The population of AGN do show a continuum of these parameters, with the CL AGN the most extreme end of the distribution.  However, these plots make it clear that the most important parameter for distinguishing CL AGN is the $R$ parameter, with CL AGN sitting in a very sparsely populated region of R values.  The log scale in these plots shows that the $R$ values of CL AGN are typically orders of magnitude higher than most AGN.  On the other hand CL AGN are in the same $\rm Max(N_{\sigma})$ range as about a third of the AGN in our sample, and are in the same $ F_{\rm \sigma, dim }$ range as about one fifth of the sample.

\subsection{ Flux  Calibration Check }
\label{sec_oiii}
Several previous studies carried out a further flux calibration based on either an [O\,{\sc iii}] ($5007$\AA) calibration or using a photometric light curve to exclude fiber position and/or aperture issues introduced from the program selection \citep{Yang2018, MacLeod2019, Green2022}. In this work, we combine both methods to further check the selected CL AGN.

The  [O\,{\sc iii}]  flux should remain relatively unchanged since the [O\,{\sc iii}] emission line probes the region reaching hundreds of parsecs to several kiloparsecs away from the central ionization source and originates from the narrow line region and host galaxy even though there are still some controversies about [O\,{\sc iii}] flux variation \citep{Zhang2016,Barth2016,Yan2019}. If the [O\,{\sc iii}]  flux difference between the two spectra is less than 15\%, the possible reasons are fiber diameter differences, changes in the observation seeing, and possible flux changes, in which case recalibration of the spectrum will cause a fake CL phenomenon. Therefore, we adopted the [O\,{\sc iii}]  calibration for those [O\,{\sc iii}] luminosities that changed between the DESI and SDSS spectra beyond the 15\% deviation.  We remove those objects that do not satisfy the criteria outlined in Section \ref{sec_definition} after calibration to ensure the accuracy of our sample.  

To align the resolution of the DESI and SDSS spectra, we employ a Gaussian kernel to convolve the DESI spectra (R $\sim$ 4000) to match the lower resolution of the SDSS spectra (R $\sim$ 2000). Following the approach outlined in \cite{Du2018}, we conduct an [O\,{\sc iii}]-based calibration to scale the flux of the SDSS spectrum, taking into account the [O\,{\sc iii}] flux ratio between DESI and SDSS for AGN at $z \leq 0.75$.

However, considering the DESI have small fiber diameter (1.5\arcsec) than SDSS fiber diameter (2.0\arcsec), the extended  O\,{\sc iii}] region (which can extend to more than 10 kpc) may all outside of the fiber radius \citep{Goddard2010M}.  This would change the [O\,{\sc iii}] flux and can be exacerbated by variable seeing. Such an effect would be more significant at lower redshifts. In these cases, [O\,{\sc iii}] calibration may enlarge the flux of  emission line or continuum,  we thus only use [O\,{\sc iii}] calibration to identify cases where the variation may be spurious, and use the non-calibrated spectra in the final results shown in Figure \ref{fig_example}.

The photometric light curve serves as a criterion for checking the presence of a flux calibration problem between the DESI and SDSS spectra. As mentioned in \cite{Guo2020}, the SDSS spectrum may exhibit problematic spectral flux calibration, resulting in scatter of up to $20\%$ at high redshift, known as fiber drop. This effect becomes more pronounced when screening for CL AGN. Therefore, we utilized the publicly available light curves from CRTS, PTF, PS1, and ZTF to compare the relevant photometry with the pseudophotometry around the spectral Modified Julian Date (MJD). It is important to note that the CRTS data were unfiltered at the time of observation and were subsequently converted to the $V$-band, which may introduce inaccuracies in the pseudophotometry. Finally, 74 CL AGN and 47 candidates were removed if their spectral photometry significantly deviated from the light curve, typically exceeding 0.5 magnitude.

 Although the pseudophotometry for J164331.90+30483 5.5, J221044.76+245958.0, and J221925.57+272806.4 shows a deviation of approximately 0.5 magnitudes compared to the ZTF light curve, we have decided to retain these objects as they still meet our criteria after calibration based on [O,{\sc iii}] measurements. Furthermore, since the redshift of these three objects is $z<0.25$, the observed deviation is primarily attributed to inconsistencies in the contribution of starlight between the spectral diameter coverage and the ZTF psf-based photometric modeling. The DESI Legacy Images confirm the presence of significant extended source features in their host galaxies, as shown in Figure \ref{fig_image}. It is worth noting that the DESI and SDSS fiber diameters capture less starlight from the host galaxy compared to the ZTF light curve.

After conducting [O\,{\sc iii}]-based calibration and analyzing the photometric light curve, we have removed 73 CL AGN (38 candidates) from the original sample of 130 CL AGN (91 candidates). Among these, two CL AGN were identified to have DESI flux calibration problems. Figure \ref{fig_problem} illustrates the SDSS and DESI fiber drop effect.

\begin{deluxetable*}{cccccccccccccc}
\tablecolumns{12}
\setcounter{table}{3}
\label{tab_CL}
\tabletypesize{\footnotesize}
\tabcaption{\centering Sample properties for the CL AGN presented in this study
}
\colnumbers
\tablehead{
\colhead{Object Name} &
\colhead{R.A.} &
\colhead{Dec.} &
\colhead{Redshift} &
\colhead{$g$(mag)} &
\colhead{$r$(mag)} &
\colhead{$z$(mag)} &
\colhead{$\rm MJD_{1}$} &
\colhead{$\rm MJD_{2}$} &
\colhead{$\rm N_{\sigma}$} &
\colhead{R} &
\colhead{$F_{\rm \sigma, \rm dim }$} &
\colhead{transition} &
\colhead{Line} 
}
\startdata
075448.10+345828.5 & 118.7004 & 34.9746 & 0.7311 & 21.49 & 21.50 & 20.92 & 55488 & 59226 & 5.22 & 2.85 & 6.70 & turn-off & $\rm H\beta$\\
085913.72+323050.8 & 134.8072 & 32.5141 & 1.1217 & 20.48 & 20.18 & 20.35 & 52989 & 59226 & 7.89 & 3.32 & 3.41 & turn-on &  C\,{\sc iii}]\\
095035.55+321601.0 & 147.6482 & 32.2670 & 1.7181 & 20.59 & 20.54 & 20.33 & 56325 & 59280 & 4.40 & 2.24 & 4.46 & turn-on &  C\,{\sc iii}]\\
 & &  & &  &  & & &  &  4.32 & 2.13 & 3.18 & turn-on & Mg\,{\sc ii}\\
103818.29+332437.2 & 159.5762 & 33.4103 & 1.2028 & 21.10 & 20.65 & 20.63 & 58497 & 59224 & 3.93 & 2.40 & 3.19 & turn-on &  C\,{\sc iii}]\\
113737.38+511839.9 & 174.4058 & 51.3111 & 1.0671 & 20.70 & 20.48 & 20.39 & 56416 & 59307 & 4.34 & 2.68 & 1.51 & turn-off &  C\,{\sc iii}]\\
115103.77+530140.6 & 177.7657 & 53.0280 & 0.5475 & 22.22 & 21.83 & 20.78 & 57423 & 59307 & 3.76 & 49.49 & 0.09 & turn-off & Mg\,{\sc ii}\\
 & &  & &  &  & & &  &  8.05 & 7.27 & 2.06 & turn-off & $\rm H\beta$\\
115210.24+520205.1 & 178.0427 & 52.0347 & 0.6179 & 21.33 & 21.36 & 20.97 & 56416 & 59307 & 4.10 & 4.53 & 2.52 & turn-on & $\rm H\beta$\\
115403.00+003154.0 & 178.5125 & 0.5317 & 0.4484 & 21.10 & 20.21 & 19.44 & 51943 & 59312 & 4.92 & 6.35 & 1.20 & turn-off & $\rm H\beta$\\
 & &  & &  &  & & &  &  5.79 & 8.28 & 1.41 & turn-off & Mg\,{\sc ii}\\
121033.30-011755.6$^{*}$ & 182.6388 & -1.2988 & 0.5434 & 20.03 & 20.04 & 19.49 & 52367 & 59314 & 14.60 & 7.46 & 5.96 & turn-off & $\rm H\beta$\\
122319.70+312737.0 & 185.8321 & 31.4603 & 1.5188 & 21.89 & 21.54 & 21.24 & 58488 & 59291 & 3.60 & 2.72 & 2.58 & turn-on & Mg\,{\sc ii}\\
125545.34+232348.0 & 193.9390 & 23.3967 & 0.5096 & 21.18 & 21.20 & 20.51 & 56312 & 59338 & 3.92 & 2.45 & 4.39 & turn-off & $\rm H\beta$\\
125610.42+260103.4 & 194.0434 & 26.0176 & 1.1957 & 20.57 & 19.89 & 19.94 & 54505 & 59337 & 3.81 & 4.80 & 2.34 & turn-on &  C\,{\sc iii}]\\
133044.43+072628.8 & 202.6851 & 7.4413 & 0.4634 & 19.48 & 19.44 & 19.11 & 53556 & 59370 & 6.15 & 2.77 & 4.84 & turn-off & Mg\,{\sc ii}\\
 & &  & &  &  & & &  &  7.89 & 3.18 & 4.90 & turn-off & $\rm H\beta$\\
133344.70+335622.7 & 203.4363 & 33.9396 & 1.1452 & 21.50 & 21.12 & 21.30 & 58492 & 59292 & 4.38 & 3.09 & 2.82 & turn-on &  C\,{\sc iii}]\\
134037.05+131819.2 & 205.1544 & 13.3054 & 0.3478 & 20.78 & 19.82 & 19.08 & 53858 & 59372 & 7.61 & 2.76 & 3.95 & turn-off & $\rm H\beta$\\
134554.00+084537.3 & 206.4750 & 8.7603 & 0.6020 & 20.76 & 20.37 & 20.21 & 55973 & 59371 & 4.27 & 3.68 & 5.28 & turn-on & Mg\,{\sc ii}\\
135613.60+001958.6 & 209.0567 & 0.3330 & 2.3286 & 21.62 & 21.45 & 21.21 & 55631 & 59376 & 4.19 & 2.58 & 4.89 & turn-off &  C\,{\sc iv}\\
135624.58+350813.1 & 209.1025 & 35.1370 & 0.7983 & 22.18 & 21.66 & 20.74 & 55268 & 59350 & 5.03 & 3.68 & 2.86 & turn-off & Mg\,{\sc ii}\\
135801.62+115331.8 & 209.5068 & 11.8922 & 0.2667 & 20.60 & 19.68 & 19.14 & 53144 & 59378 & 10.86 & 6.76 & 4.45 & turn-off & $\rm H\beta$\\
140659.07+032601.9 & 211.7461 & 3.4339 & 0.5995 & 19.94 & 19.86 & 19.57 & 52045 & 59377 & 6.32 & 2.24 & 7.09 & turn-off & $\rm H\beta$\\
140957.72-012850.5 & 212.4905 & -1.4807 & 0.1352 & 19.05 & 18.21 & 17.49 & 55383 & 59338 & 10.18 & 2.47 & 6.23 & turn-off & $\rm H\beta$\\
141535.46+022338.7 & 213.8977 & 2.3941 & 0.3519 & 20.50 & 19.49 & 18.60 & 51994 & 59370 & 16.25 & 9.36 & 14.27 & turn-off & $\rm H\alpha$\\
 & &  & &  &  & & &  &  16.23 & 15.84 & 3.64 & turn-off & $\rm H\beta$\\
141801.50+525200.7 & 214.5063 & 52.8669 & 1.1555 & 21.98 & 21.50 & 21.02 & 56755 & 59314 & 3.27 & 5.63 & 1.82 & turn-off &  C\,{\sc iii}]\\
141923.44-030458.7 & 214.8477 & -3.0830 & 2.6846 & 20.47 & 20.39 & 20.34 & 55333 & 59368 & 4.52 & 3.18 & 1.94 & turn-on &  C\,{\sc iii}]\\
142641.03+521214.3 & 216.6710 & 52.2040 & 1.0493 & 21.48 & 21.19 & 20.94 & 56448 & 59260 & 3.70 & 70.89 & 0.04 & turn-on &  C\,{\sc iii}]\\
144003.98+061936.5 & 220.0166 & 6.3268 & 0.5395 & 19.44 & 19.31 & 18.96 & 55684 & 59377 & 8.43 & 2.54 & 6.20 & turn-on & $\rm H\beta$\\
144051.17+024415.8 & 220.2132 & 2.7377 & 0.3963 & 20.09 & 19.77 & 19.10 & 55620 & 59379 & 7.56 & 5.08 & 2.68 & turn-off & $\rm H\beta$\\
145913.90+360051.4 & 224.8079 & 36.0143 & 0.5526 & 20.52 & 20.10 & 20.08 & 53121 & 59351 & 8.26 & 3.67 & 3.89 & turn-on & $\rm H\beta$\\
152517.57+401357.6 & 231.3232 & 40.2327 & 0.3838 & 19.99 & 19.77 & 19.24 & 58175 & 59354 & 7.38 & 5.25 & 3.75 & turn-off & $\rm H\beta$\\
152551.37+184552.0 & 231.4640 & 18.7645 & 0.4360 & 19.59 & 18.84 & 18.47 & 54328 & 59370 & 8.42 & 15.78 & 1.06 & turn-off & $\rm H\beta$\\
152908.75+083203.5 & 232.2865 & 8.5343 & 2.6581 & 21.03 & 20.76 & 20.57 & 56002 & 59366 & 4.55 & 4.29 & 1.91 & turn-on &  C\,{\sc iii}]\\
153149.94+372755.4 & 232.9581 & 37.4654 & 0.9951 & 20.94 & 20.85 & 20.81 & 58230 & 59351 & 3.76 & 1.92 & 3.37 & turn-off &  C\,{\sc iii}]\\
153714.41+454347.7 & 234.3101 & 45.7299 & 0.4688 & 21.09 & 20.24 & 19.28 & 52781 & 59385 & 8.87 & 3.40 & 5.64 & turn-off & Mg\,{\sc ii}\\
 & &  & &  &  & & &  &  9.28 & 12.18 & 1.31 & turn-off & $\rm H\beta$\\
153938.92+373853.8 & 234.9122 & 37.6483 & 1.4187 & 22.57 & 21.90 & 21.91 & 58258 & 59375 & 3.71 & 2.16 & 3.74 & turn-on &  C\,{\sc iii}]\\
154742.72+012541.0 & 236.9280 & 1.4281 & 0.6879 & 19.86 & 19.88 & 19.66 & 55359 & 59352 & 14.93 & 3.10 & 8.47 & turn-on & $\rm H\beta$\\
155426.13+200527.7 & 238.6089 & 20.0910 & 0.5215 & 22.21 & 20.94 & 20.09 & 55332 & 59374 & 6.71 & 11.48 & 0.89 & turn-off & Mg\,{\sc ii}\\
 & &  & &  &  & & &  &  5.42 & 105.42 & 0.11 & turn-off & $\rm H\beta$\\
160047.63+331310.7 & 240.1985 & 33.2196 & 0.3608 & 20.01 & 19.35 & 18.59 & 53142 & 59358 & 10.67 & 2.11 & 6.63 & turn-off & $\rm H\beta$\\
160139.10+412529.3 & 240.4130 & 41.4248 & 0.4645 & 21.35 & 20.56 & 19.58 & 52824 & 59354 & 9.89 & 3.01 & 4.57 & turn-off & $\rm H\beta$\\
160310.99+432928.9 & 240.7958 & 43.4914 & 0.4923 & 20.33 & 20.03 & 19.35 & 52756 & 59321 & 8.37 & 5.21 & 8.68 & turn-off & $\rm H\beta$\\
160337.12+242513.9 & 240.9047 & 24.4205 & 0.7058 & 22.13 & 22.14 & 21.17 & 55327 & 59376 & 9.62 & 2.89 & 9.28 & turn-off & $\rm H\beta$\\
160730.20+560305.5 & 241.8759 & 56.0515 & 0.7172 & 21.62 & 21.35 & 20.70 & 56430 & 59311 & 6.57 & 2.61 & 3.59 & turn-on & $\rm H\beta$\\
161542.41+233143.9 & 243.9267 & 23.5289 & 0.5757 & 20.48 & 20.10 & 19.67 & 55331 & 59376 & 7.36 & 2.41 & 4.31 & turn-off & $\rm H\beta$\\
161903.04+540529.0 & 244.7627 & 54.0914 & 0.6033 & 20.76 & 20.78 & 20.43 & 58248 & 59311 & 5.09 & 4.24 & 2.65 & turn-on & $\rm H\beta$\\
162106.25+371950.7 & 245.2761 & 37.3307 & 1.3883 & 21.28 & 20.89 & 20.96 & 58247 & 59350 & 3.84 & 1.83 & 2.16 & turn-off &  C\,{\sc iii}]\\
 162829.18+432948.5$^{*}$ & 247.1216 & 43.4968 & 0.2599 & 19.80 & 19.22 & 18.73 & 52057 & 59316 & 9.53 & 4.22 & 8.11 & turn-off & $\rm H\beta$\\
164331.90+304835.5 & 250.8830 & 30.8099 & 0.1837 & 19.71 & 18.75 & 18.09 & 52793 & 59386 & 12.69 & 35.76 & 0.67 & turn-off & $\rm H\beta$\\
164709.87+532202.2 & 251.7911 & 53.3673 & 1.2008 & 21.45 & 21.03 & 20.93 & 57190 & 59383 & 7.06 & 3.72 & 4.89 & turn-on &  C\,{\sc iii}]\\
164725.15+351754.3 & 251.8548 & 35.2985 & 0.6886 & 20.95 & 20.91 & 20.82 & 55828 & 59312 & 4.64 & 12.35 & 0.71 & turn-on & $\rm H\beta$ 
\enddata
\end{deluxetable*}

\begin{deluxetable*}{cccccccccccccc}
\tablecolumns{12}
\setcounter{table}{3}
\tabletypesize{\footnotesize}
\tabcaption{\centering Sample properties for the CL AGN presented in this study
}
\colnumbers
\tablehead{
\colhead{Object Name} &
\colhead{R.A.} &
\colhead{Dec.} &
\colhead{Redshift} &
\colhead{$g$(mag)} &
\colhead{$r$(mag)} &
\colhead{$z$(mag)} &
\colhead{$\rm MJD_{1}$} &
\colhead{$\rm MJD_{2}$} &
\colhead{$\rm N_{\sigma}$} &
\colhead{R} &
\colhead{$F_{\rm \sigma, \rm dim }$} &
\colhead{transition} &
\colhead{Line} 
}
\startdata
164900.95+452016.8 & 252.2540 & 45.3380 & 0.5806 & 20.50 & 20.48 & 20.18 & 58021 & 59354 & 6.49 & 2.03 & 5.12 & turn-on & $\rm H\beta$\\
213135.84+001517.0 & 322.8994 & 0.2547 & 2.4755 & 21.68 & 21.39 & 20.79 & 55450 & 59382 & 4.95 & 3.07 & 2.59 & turn-on &  C\,{\sc iii}]\\
213400.68+013828.4 & 323.5029 & 1.6412 & 0.9973 & 20.87 & 20.61 & 20.52 & 58040 & 59380 & 4.20 & 20.09 & 0.56 & turn-on &  C\,{\sc iii}]\\
213628.50-003811.8 & 324.1188 & -0.6366 & 2.2368 & 21.94 & 22.02 & 21.34 & 55450 & 59382 & 4.89 & 4.17 & 2.35 & turn-off &  C\,{\sc iii}]\\
221044.76+245958.0 & 332.6865 & 24.9995 & 0.1199 & 20.14 & 19.02 & 18.19 & 56213 & 59392 & 3.67 & 5.33 & 5.52 & turn-on & $\rm H\beta$\\
 & &  & &  &  & & &  &  3.05 & 1.82 & 55.05 & turn-on & $\rm H\alpha$\\
221925.57+272806.4 & 334.8566 & 27.4685 & 0.2282 & 20.60 & 19.58 & 18.84 & 57575 & 59392 & 19.17 & 3.14 & 5.57 & turn-off & $\rm H\beta$\\
223440.56+272437.9 & 338.6690 & 27.4105 & 1.0203 & 21.00 & 20.72 & 20.68 & 57654 & 59403 & 4.03 & 7.81 & 0.66 & turn-on &  C\,{\sc iii}]\\
224657.70-003242.5 & 341.7404 & -0.5451 & 0.4157 & 20.43 & 19.75 & 18.98 & 52590 & 59404 & 4.22 & 3.38 & 6.68 & turn-on & $\rm H\beta$
\enddata
\tablecomments{Columns: (1) name, (2) right ascension,  (3) declination, (4) redshift, (5) photometric magnitude for $g$ band, (6) photometric magnitude for $r$ band, (7) photometric magnitude for $z$ band, (8) MJD for SDSS spectrum, (9) MJD for DESI spectrum, (10)  flux deviation from Equation (\ref{eq_N}), (11) flux change ratio from Equation (\ref{eq_R}), (12) the significance of the Broad Emission Line,  (13) changing-look type,  (14) target Broad Emission Line. \\
We note that he three asterisk targets were replicated from other studies. (121033.30-011755.6  refers to \cite{MacLeod2019}, 162829.18+432948.5 refers to \cite{Zeltyn2022})}
\end{deluxetable*}

\vspace{-20mm}

\begin{deluxetable*}{cccccccccccccc}
\label{tab_cl_ca}

\tablecolumns{12}
\tabletypesize{\footnotesize}
\tabcaption{\centering Sample properties for CL candidate features in confirmed CL AGN. 
}
\colnumbers
\tablehead{
\colhead{Object Name} &
\colhead{R.A.} &
\colhead{Dec.} &
\colhead{Redshift} &
\colhead{$g$(mag)} &
\colhead{$r$(mag)} &
\colhead{$z$(mag)} &
\colhead{$\rm MJD_{1}$} &
\colhead{$\rm MJD_{2}$} &
\colhead{$\rm N_{\sigma}$} &
\colhead{R} &
\colhead{$F_{\rm \sigma, \rm dim }$} &
\colhead{transition} &
\colhead{Line} 
}
\startdata
075448.10+345828.5 & 118.7004 & 34.9746 & 0.7311 & 21.49 & 21.50 & 20.92 & 55488 & 59226 & 14.52 & 2.20 & 16.65 & turn-off & Mg\,{\sc ii}\\
121033.30-011755.6 & 182.6388 & -1.2988 & 0.5434 & 20.03 & 20.04 & 19.49 & 52367 & 59314 & 13.96 & 1.62 & 18.31 & turn-off & Mg\,{\sc ii}\\
134037.05+131819.2 & 205.1544 & 13.3054 & 0.3478 & 20.78 & 19.82 & 19.08 & 53858 & 59372 & 10.49 & 2.02 & 18.05 & turn-off & $\rm H\alpha$\\
144051.17+024415.8 & 220.2132 & 2.7377 & 0.3963 & 20.09 & 19.77 & 19.10 & 55620 & 59379 & 9.47 & 1.61 & 14.16 & turn-off & $\rm H\alpha$\\
152517.57+401357.6 & 231.3232 & 40.2327 & 0.3838 & 19.99 & 19.77 & 19.24 & 58175 & 59354 & 12.50 & 2.61 & 31.46 & turn-off & $\rm H\alpha$\\
 & &  & &  &  & & &  &  7.87 & 2.29 & 11.76 & turn-off & Mg\,{\sc ii}\\
160139.10+412529.3 & 240.4130 & 41.4248 & 0.4645 & 21.35 & 20.56 & 19.58 & 52824 & 59354 & 10.53 & 2.76 & 12.60 & turn-off & Mg\,{\sc ii}\\
160310.99+432928.9 & 240.7958 & 43.4914 & 0.4923 & 20.33 & 20.03 & 19.35 & 52756 & 59321 & 8.42 & 1.59 & 20.09 & turn-off & Mg\,{\sc ii}\\
162829.18+432948.5 & 247.1216 & 43.4968 & 0.2599 & 19.80 & 19.22 & 18.73 & 52057 & 59316 & 13.43 & 1.93 & 63.19 & turn-off & $\rm H\alpha$\\
164331.90+304835.5 & 250.8830 & 30.8099 & 0.1837 & 19.71 & 18.75 & 18.09 & 52793 & 59386 & 25.42 & 1.81 & 73.12 & turn-off & $\rm H\alpha$\\
\enddata
\tablecomments{The columns are same as Table \ref{tab_CL}}
\end{deluxetable*}

\begin{figure*}
\centering
\includegraphics[width=0.75\textwidth]{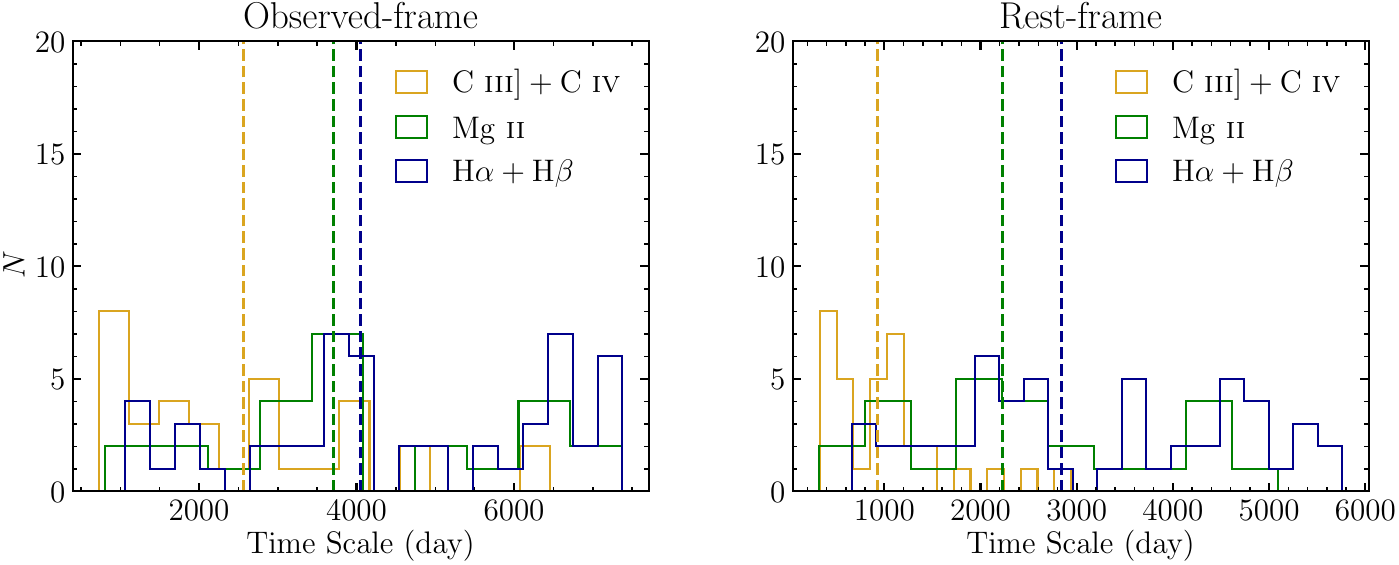}
\caption{The histograms of the upper limit of the transition time scale for CL AGN  and CL candidates in  observed\mbox{-}frame (left panel) and  rest\mbox{-}frame (right panel). The dash lines represent the median value of three samples. While it seems like C\,{\sc iii}] and C\,{\sc iv} have lower rest-frame time scales than the other lines, the observed-frame histogram shows that that is partially due to a selection effect, since time-dilation has made it impossible from our sample to detect carbon rest-frame lags longer than 3500 days.}
\label{fig_timescale}
\end{figure*}

\begin{figure*}
\centering
\includegraphics[width=1\textwidth]{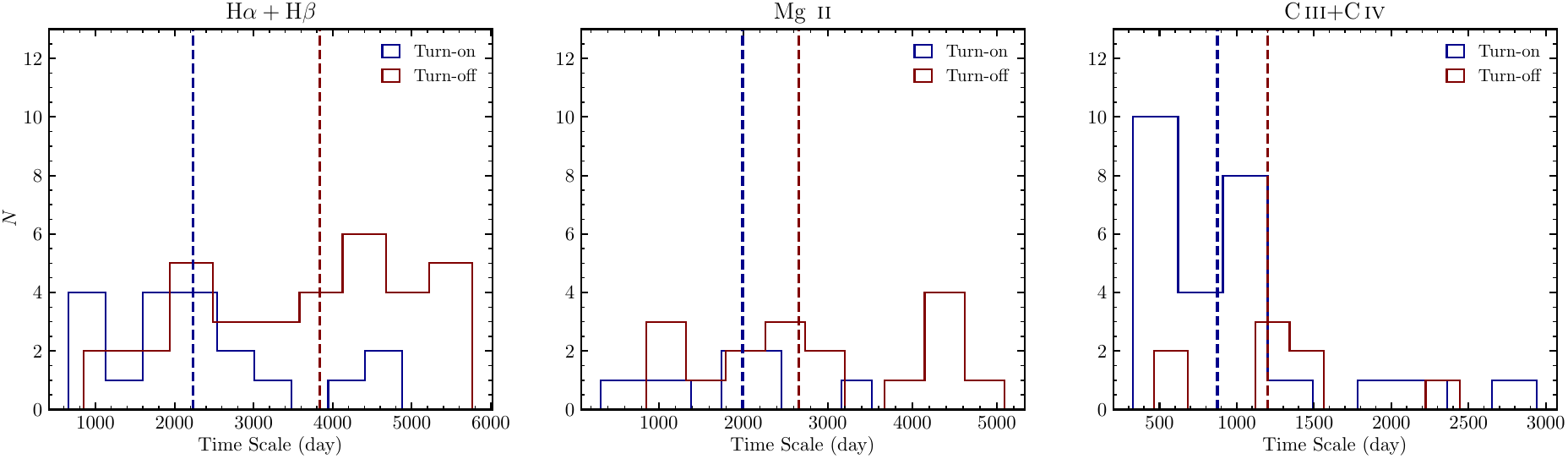}
\caption{The histograms of the upper limit of the transition time scale for $\rm{H}\alpha+ \rm{H}\beta$ (left panel),  Mg\,{\sc ii}(middle panel),  C\,{\sc iii}] +C\,{\sc iv} (right panel)  CL AGN and CL candidates in  rest\mbox{-}frame (right panel). The dash lines represent the median value of samples.}
\label{fig_timescale2}
\end{figure*}

\section{Results}
\label{sec_results}
\subsection{CL AGN and Candidates }

Within the current DESI project, we obtain 56  CL AGN based on the variability definition outlined in Section \ref{sec_definition} and VI in Section \ref{sec_VI}, which are listed in Table \ref{tab1} and \ref{tab2}. Of these 56  AGN,  J162829.18+432948.5 was previously discovered in \cite{Zeltyn2022} and   121033.30-011755.6  was selected as CL candidates reported by \cite{MacLeod2019} to have high optical variability, although were lacking spectra.  Specifically, we identify  2 $\rm{H}\alpha$,  34 $\rm{H}\beta$, 9  Mg\,{\sc ii}, 18  C\,{\sc iii}] and 1  C\,{\sc iv} CL objects. While J221044.76+245958.0 is classified as CL candidates in the selection criteria, we moved it from  CL candidates into CL AGN by VI because the  significance of their broad components are increased by narrow lines.
Figure \ref{fig_example} shows five examples of $\rm{H}\alpha,  \rm{H}\beta$, Mg\,{\sc ii},  C\,{\sc iii}], and  C\,{\sc iv} CL AGN respectively, where the difference spectrum is the subtraction of the dim spectrum from the bright spectrum.  Table \ref{tab_CL} summarizes the properties of the 57  CL AGN.

To further study the whole process of CL AGN, we also search for CL AGN candidates which are in the process of changing or about to happen. In the 56 CL AGN we have discovered, 9 AGN have candidate additional BEL variations (listed in Table \ref{tab_cl_ca}). We also provide  another 44 CL candidates for five BELs as shown in Table \ref{tab_CL2}. The whole CL  candidate sample contains 53 AGN, including 7 $\rm{H}\alpha$, 17 $\rm{H}\beta$, 16 Mg\,{\sc ii}, 10 C\,{\sc iii}] and 5 C\,{\sc iv} CL candidates. To highlight the difference between the confirmed CL AGN and the candidates, Figure \ref{fig_example4} provides five selected examples of CL candidate spectra. Figure \ref{fig_example5} shows the rest 50 CL AGN  while the full CL AGN candidate figures will be made available in its entirety in machine-readable form.

\subsection{ Time Scale and Transition State} 
\label{sec_timescale}

In Figure \ref{fig_timescale}, we plot the time scale of both the 56 confirmed CL AGN and the 44 CL candidates between the two observations in the observed\mbox{-}frame and rest\mbox{-}frame. We note that the  number of CL AGN that show different BEL behavior from the same source are counted repeatedly. For instance, a case with varying $\rm{H}\beta$ and Mg\,{\sc ii} will be once in the Mg\,{\sc ii} sample (green  histogram) and once in the $\rm{H}\beta$ sample (blue histogram). \cite{Zeltyn2022} reported 
a time scale for the $\rm{H}\beta$ variation observed in J162829.18+432948.5 of only a few months (73 days), whereas we probe a time scale of up to 5643 days between the SDSS and DESI observations.  The time scale represents the upper limit of the transition time scale and ranges from 330 to 5762 days.  
While the observed-frame histogram suggests that some differences may be attributed to a selection effect or observational bias, systematic differences between different samples persist. One reason may be due to the selection bias of  a small sample size, which requires further investigation with a larger sample.

In our sample of 56 CL AGN, 24 AGN exhibit turn-on behavior while 32 AGN exhibit turn-off behavior, as listed in Table \ref{tab_CL}.  The proportion of turn-off and turn-on for  CL AGN  is consistent with previous work \citep{MacLeod2016, Yang2018, MacLeod2019, Green2022, WangJ2022}. The redshift and magnitude distributions for  turn-off and turn-on CL AGN are shown in  the bottom panles of  Figure \ref{fig_redshift}. Figure \ref{fig_timescale2} show the timescale for  different emission lines with transition state, from which we can see that the turn-off time scales are larger than  turn-on time scales in all  emission lines. However, due to the limited number of samples, whether there is a different time scales of turn-on and turn-off CL AGN still needs further more complete sample verification.

\subsection{Physical Origin }
\label{sec_origin}

Despite the limitations in information about the continuum luminosity and power-law index, it is evident that the majority of CL AGN from Figure \ref{fig_example}, particularly for $\rm{H}\alpha$ and $\rm{H}\beta$, display the characteristic of being bluer when brighter (the optical and ultraviolet emission from the nucleus of an AGN becomes bluer when the overall brightness of the AGN increases), which is in line with many previous studies \citep{LaMassa2015, Runnoe2016, MacLeod2016, Gezari2017, Yang2018, Graham2020, Green2022}. From a quantitative standpoint, \cite{MacLeod2019} and \cite{Green2022}  presented two separate CL AGN samples obtained through the method of spectral fitting, and both find a strong correlation between the change in $\rm{H}\beta$ flux and the variation in continuum flux (refer to Figure 3 in \cite{MacLeod2019} and Figure 8 in \cite{Green2022} for more details). This close relationship between the continuum and BEL variation suggests that the radiation from the accretion disk continues to drive the evolution of the broad line region even during the changing state. This phenomenon also implies that there is a rapid exchange of both material and energy occurring between the accretion disk and the broad line region, indicating that the origin of CL AGN is more likely to come from changes in the accretion rate.

Despite this, there are still some uncertainties in the identification of CL AGN. According to \citealt{MacLeod2016} and \citealt{Green2022}, some CL AGN exhibit back-and-forth variability, where the weak $\rm{H}\beta$ broad component disappears in one epoch but reappears in another. This suggests that the discovery of CL AGN is like a random sampling from a pool with constantly changing BELs.

One possible explanation for the CL behavior is that the AGN is undergoing a transition from a low-to-high accretion rate state or vice versa. In the low-accretion rate state, the emission from the central region is relatively weak, while in the high-accretion rate state, the emission is much stronger. The transition between these two states can be triggered by various physical processes, such as the instability of the accretion disk or changes in the mass-supply rate. \cite{Elitzur2009} proposed that AGN may have a critical luminosity that switches the BEL appearance and disappearance. This model is also supported by the discovered CL AGN, which shows a low accretion rate and low Eddington ratio (usually $< 0.1$) in the dim spectrum. Given that the radiation regions of C\,{\sc iii}] and C\,{\sc iv} are relatively small, another potential scenario that could describe the C\,{\sc iii}] and C\,{\sc iv} CL behavior is the association with an outflow

Given that the majority of CL AGN display the ``bluer when brighter" trend, another potential explanation could be a change in obscuration of the Broad Line Region, either due to dust clouds that move in and out of our line-of-sight or powerful winds that ``blow-out'' the dust surrounding the nucleus.  However, the former scenario has not been supported  by polarization observations, time scale calculations, and mid-infrared features for the majority  CL AGN \citep{LaMassa2015, MacLeod2016, Sheng2017}. 
Many of the turn-on CL AGN exhibit a wavelength-dependent gain in continuum flux, appearing to change from a ``red quasar’’ (e.g., \citealt{Klindt2019,Fawcett2020}) to a typical blue quasar (see Figure \ref{fig_example}).  This could indicate that dust is being removed from these systems, which may be evidence that these AGN are undergoing a ``blow-out’’ phase, in which powerful outflows clear out the surrounding gas/dust \citep{Glikman2017,Stacey2022,Fawcett2022}.  Using spectroscopy over a 19-year period, \cite{Yi2022} discovered a turn-on CL AGN that hosted powerful outflows, concluding that the quasar was shedding a surrounding dust cocoon, transitioning to a blue quasar. However, it is still unclear whether changes in the accretion disc could account for the change in continuum flux and also whether the difference in the dust extinction values over a 10–20-year timescale is consistent with a blow-out phase.   Several possible models have also been proposed to explain the CL phenomenon, including the cooling front model caused by a change in the magnetic field \citep{Ross2020}, close binary black hole model \citep{Wangjianmin2020}, magnetic field-coupled accretion outflow model \citep{Feng2021}.
In the picture of an advection-dominated accretion flow model, CL behavior might be caused by an intermittent accretion stream, which would result in a back-and-forth BEL pattern in the AGN \citep{Noda2018}. This study is  first step in an ongoing effort to constrain or test these models by analyzing a large sample of CL AGN in the  DESI project.


\section{Conclusion}
\label{sec_summary}
In DESI early data,  we carry out a systematic search for CL AGN through cross-matching with the SDSS DR16 spectroscopic database. From a parent sample of 82,653 AGN, the following summarizes our main findings:
\begin{itemize}
    \item  We have compiled a sample of 56 CL AGN selected based on VI and our defined criteria, which includes 2 $\rm{H}\alpha$, 34 $\rm{H}\beta$, 8 Mg\,{\sc ii}, 18 C\,{\sc iii}], and 1 C\,{\sc iv} CL behaviors. 
    \item  We find 10 candidate CL features in confirmed CL AGN and provide another 44 CL AGN candidates, which show the dramatic flux variation of  emission lines but remain a significant broad component. 

    \item We identify 32 CL AGN that display a turn-off transition and 24 CL AGN as a turn-on transition, whose time scales show significant differences, with turn-off larger than turn-on.
    
    \item We confirm the tendency of bluer when brighter, which is consistent with the behavior of previously discovered CL AGN. 
\end{itemize}
Although our current research focuses on compiling a substantial sample of CL AGN, future spectral decomposition work (such as, a more accurate  black hole mass and accretion rate, a detail broad emission line measurement in the  dim state) and dust extinction tests that will analyze the broad emission lines and continuum could greatly enhance our understanding of the physical mechanisms behind CL activity \citep{Fawcett2020, Fawcett2022}.   In addition, future studies of CL AGN will be crucial in advancing our understanding of the growth and evolution of supermassive black holes and the properties of the circumnuclear material in AGN \citep{Ho2005}. The ongoing DESI project provides an exciting opportunity to gain further insight into the CL physical mechanism. In our follow-up work, we would attempt to verify the CL sequence and conduct further spectral decomposition to analyze the physical mechanism \citep{MacLeod2016,guohengxiao2019,Green2022}. Additionally, we will continue to search for CL AGN in DESI DR1 and investigate the difference and relationship between CL AGN and normal AGN.


\begin{deluxetable*}{cccccccccccccc}
\tablecolumns{12}
\label{tab_CL2}
\tabletypesize{\footnotesize}
\tabcaption{\centering Sample properties for the CL AGN candidates presented in this study
}
\colnumbers
\tablehead{
\colhead{Object Name} &
\colhead{R.A.} &
\colhead{Dec.} &
\colhead{Redshift} &
\colhead{$g$(mag)} &
\colhead{$r$(mag)} &
\colhead{$z$(mag)} &
\colhead{$\rm MJD_{1}$} &
\colhead{$\rm MJD_{2}$} &
\colhead{$\rm N_{\sigma}$} &
\colhead{R} &
\colhead{$F_{\rm \sigma, \rm dim }$} &
\colhead{transition} &
\colhead{Line}
}
\startdata
091452.90+323347.1 & 138.7204 & 32.5631 & 1.5782 & 20.99 & 20.88 & 20.63 & 54550 & 59251 & 6.89 & 1.73 & 5.54 & turn-on &  C\,{\sc iii}]\\
102012.84+324737.2 & 155.0535 & 32.7937 & 0.6200 & 20.20 & 20.13 & 19.82 & 53442 & 59218 & 7.59 & 1.90 & 35.65 & turn-off & $\rm H\beta$\\
102104.98+440355.5 & 155.2708 & 44.0654 & 0.6046 & 20.12 & 20.06 & 19.70 & 57419 & 59218 & 4.13 & 1.74 & 6.37 & turn-on & $\rm H\beta$\\
111634.91+540138.8 & 169.1455 & 54.0274 & 1.0618 & 21.44 & 20.90 & 20.73 & 57135 & 59307 & 5.49 & 1.66 & 4.98 & turn-off & Mg\,{\sc ii}\\
111938.02+513315.5 & 169.9084 & 51.5543 & 0.1071 & 18.35 & 18.46 & 17.91 & 57071 & 59298 & 33.71 & 3.40 & 63.43 & turn-on & $\rm H\beta$\\
112634.33+511554.5 & 171.6431 & 51.2652 & 1.5319 & 20.97 & 20.62 & 20.82 & 57071 & 59308 & 5.24 & 2.25 & 5.17 & turn-on &  C\,{\sc iii}]\\
120332.06+563100.3 & 180.8836 & 56.5168 & 1.5200 & 21.16 & 20.31 & 20.35 & 56429 & 59309 & 4.36 & 1.64 & 6.91 & turn-on &  C\,{\sc iii}]\\
123557.86+582122.9 & 188.9911 & 58.3564 & 0.2116 & 19.39 & 18.79 & 18.47 & 52790 & 59298 & 9.68 & 1.74 & 26.90 & turn-off & $\rm H\beta$\\
124446.47+591510.8 & 191.1936 & 59.2530 & 0.6836 & 20.90 & 20.98 & 20.52 & 56443 & 59308 & 7.35 & 2.03 & 7.43 & turn-off & Mg\,{\sc ii}\\
124931.53+364816.4 & 192.3814 & 36.8045 & 1.4279 & 20.73 & 20.26 & 20.43 & 57481 & 59351 & 6.08 & 1.64 & 7.68 & turn-on &  C\,{\sc iii}]\\
 & &  & &  &  & & &  &  9.16 & 1.80 & 9.89 & turn-on & Mg\,{\sc ii}\\
125646.90+233854.8 & 194.1954 & 23.6486 & 0.8159 & 20.45 & 20.54 & 20.23 & 54212 & 59338 & 11.70 & 1.94 & 28.28 & turn-off & Mg\,{\sc ii}\\
140337.55+043126.2 & 210.9065 & 4.5239 & 0.5378 & 19.79 & 19.77 & 19.85 & 52045 & 59312 & 11.43 & 2.32 & 18.06 & turn-off & $\rm H\beta$\\
141735.11+043954.6 & 214.3963 & 4.6652 & 0.4371 & 21.38 & 20.61 & 19.82 & 55654 & 59369 & 3.92 & 1.67 & 6.39 & turn-on & $\rm H\alpha$\\
141841.39+333245.8 & 214.6725 & 33.5460 & 0.5890 & 22.10 & 21.58 & 20.50 & 55280 & 59292 & 6.05 & 1.76 & 7.60 & turn-on & $\rm H\beta$\\
142118.41+505945.2 & 215.3268 & 50.9959 & 1.0279 & 19.87 & 19.62 & 19.77 & 56415 & 59312 & 10.19 & 2.59 & 8.54 & turn-off &  C\,{\sc iii}]\\
143421.56+044137.2 & 218.5899 & 4.6937 & 0.9070 & 21.17 & 20.96 & 20.11 & 55682 & 59393 & 5.20 & 2.04 & 11.84 & turn-on & Mg\,{\sc ii}\\
143528.95+134705.7 & 218.8706 & 13.7849 & 0.7701 & 20.47 & 20.53 & 20.01 & 56014 & 59292 & 9.79 & 1.79 & 16.35 & turn-off & $\rm H\beta$\\
144813.63+080734.2 & 222.0568 & 8.1262 & 0.4414 & 18.79 & 18.74 & 18.50 & 54555 & 59379 & 14.48 & 2.33 & 20.10 & turn-on & $\rm H\beta$\\
150232.97+062337.6 & 225.6374 & 6.3938 & 0.6430 & 19.79 & 19.88 & 19.51 & 55712 & 59377 & 7.03 & 1.96 & 6.78 & turn-on & $\rm H\beta$\\
150754.87+274718.7 & 226.9786 & 27.7885 & 3.3013 & 21.46 & 20.93 & 20.94 & 55365 & 59379 & 4.57 & 1.76 & 4.03 & turn-on &  C\,{\sc iv}\\
151859.62+061840.5 & 229.7484 & 6.3112 & 0.5505 & 21.15 & 20.70 & 20.01 & 55710 & 59368 & 7.11 & 1.70 & 8.83 & turn-off & Mg\,{\sc ii}\\
152425.40+232814.7 & 231.1058 & 23.4707 & 0.6524 & 21.90 & 22.00 & 21.63 & 55680 & 59369 & 4.33 & 1.91 & 16.60 & turn-on & Mg\,{\sc ii}\\
153644.03+330721.1 & 234.1835 & 33.1225 & 0.7503 & 20.02 & 19.91 & 19.77 & 53149 & 59309 & 7.76 & 2.13 & 4.71 & turn-on & Mg\,{\sc ii}\\
153920.83+020857.2 & 234.8368 & 2.1493 & 0.3114 & 18.99 & 18.79 & 18.23 & 54562 & 59369 & 16.76 & 1.91 & 21.26 & turn-off & $\rm H\beta$\\
153952.21+334930.8 & 234.9676 & 33.8252 & 0.3289 & 18.25 & 18.06 & 17.57 & 52823 & 59309 & 15.87 & 1.60 & 27.67 & turn-on & $\rm H\beta$\\
154025.23+211445.6 & 235.1052 & 21.2460 & 0.3907 & 19.51 & 19.19 & 18.79 & 54232 & 59376 & 18.09 & 1.63 & 42.45 & turn-off & $\rm H\beta$\\
155732.71+402546.3 & 239.3863 & 40.4295 & 1.0899 & 20.64 & 20.40 & 20.52 & 58280 & 59354 & 3.73 & 1.67 & 7.21 & turn-on &  C\,{\sc iii}]\\
160129.75+401959.5 & 240.3740 & 40.3332 & 1.5823 & 22.18 & 21.58 & 21.45 & 58280 & 59354 & 4.47 & 2.12 & 8.74 & turn-on &  C\,{\sc iv}\\
160451.29+553223.4 & 241.2137 & 55.5399 & 0.3127 & 20.92 & 19.70 & 18.93 & 57897 & 59321 & 5.21 & 1.54 & 14.11 & turn-on & $\rm H\beta$\\
160712.23+151432.0 & 241.8010 & 15.2422 & 0.7789 & 21.58 & 21.16 & 20.67 & 55660 & 59292 & 4.08 & 1.89 & 13.91 & turn-off & $\rm H\beta$\\
160808.47+093715.5 & 242.0353 & 9.6210 & 0.5624 & 21.58 & 20.72 & 19.60 & 54582 & 59368 & 4.43 & 1.60 & 9.91 & turn-off & Mg\,{\sc ii}\\
161235.23+053606.8 & 243.1468 & 5.6019 & 0.7869 & 22.21 & 21.73 & 20.45 & 55708 & 59353 & 4.80 & 1.85 & 8.73 & turn-on & Mg\,{\sc ii}\\
161812.85+294416.6 & 244.5536 & 29.7379 & 0.3125 & 20.29 & 19.68 & 18.94 & 52886 & 59358 & 13.03 & 2.18 & 36.64 & turn-off & $\rm H\alpha$\\
163959.17+511930.9 & 249.9966 & 51.3252 & 1.1318 & 20.02 & 19.73 & 20.01 & 57190 & 59383 & 7.37 & 2.14 & 14.47 & turn-on &  C\,{\sc iii}]\\
164621.95+393623.8 & 251.5915 & 39.6066 & 0.3448 & 18.86 & 18.73 & 18.25 & 52050 & 59379 & 11.75 & 1.56 & 22.80 & turn-off & $\rm H\beta$\\
165919.33+304347.0 & 254.8306 & 30.7297 & 1.0375 & 20.26 & 19.79 & 19.82 & 58526 & 59356 & 6.46 & 2.36 & 7.33 & turn-on &  C\,{\sc iii}]\\
170407.13+404747.1 & 256.0298 & 40.7964 & 1.2939 & 21.13 & 20.58 & 20.61 & 58281 & 59379 & 6.10 & 2.25 & 6.64 & turn-on &  C\,{\sc iii}]\\
170624.94+423435.1 & 256.6039 & 42.5764 & 1.5936 & 21.25 & 20.91 & 20.78 & 58281 & 59379 & 10.55 & 2.03 & 32.09 & turn-on &  C\,{\sc iv}\\
172541.38+322937.8 & 261.4224 & 32.4938 & 0.6621 & 20.59 & 20.55 & 20.41 & 55721 & 59350 & 9.98 & 2.64 & 10.07 & turn-off & $\rm H\beta$\\
205407.92+005400.9 & 313.5330 & 0.9003 & 1.6435 & 21.16 & 20.98 & 20.62 & 52914 & 59370 & 9.00 & 1.90 & 13.81 & turn-off &  C\,{\sc iii}]\\
212216.44-014959.8 & 320.5685 & -1.8333 & 0.1362 & 20.23 & 19.43 & 18.81 & 58039 & 59381 & 6.12 & 2.24 & 16.71 & turn-on & $\rm H\beta$\\
212338.71+004107.0 & 320.9113 & 0.6853 & 2.0709 & 22.63 & 22.54 & 21.89 & 57691 & 59383 & 4.48 & 1.57 & 7.63 & turn-on &  C\,{\sc iv}\\
213430.72-003906.7 & 323.6280 & -0.6519 & 1.8412 & 20.99 & 20.81 & 20.49 & 57693 & 59382 & 5.42 & 2.30 & 8.13 & turn-on &  C\,{\sc iv}\\
221932.80+251850.4 & 334.8867 & 25.3140 & 0.6264 & 22.10 & 21.37 & 20.40 & 56210 & 59403 & 3.83 & 1.95 & 8.88 & turn-off & Mg\,{\sc ii}
 \enddata
\tablecomments{The columns are same as Table \ref{tab_CL}}
\end{deluxetable*}

\vspace{-8mm}
\section*{acknowledgements}
We really appreciate the referee for useful and precious comments that significantly improved the manuscript. Wei-Jian Guo would thank Suijian Xue, Luis C. Ho and Jian-Min Wang for valuable discussions and comments. We also thank DESI Publication Board Hander (Antonella Palmese) to provide timely help. This work is supported by the National Key R\&D Program of China (grant No. 2022YFA1602902), National Natural Science Foundation of China (NSFC; grant Nos. 12120101003, 12373010, and 11890691), and Beijing Municipal Natural Science Foundation (grant No. 1222028). We acknowledge the science research grants from the China Manned Space Project with Nos. CMS-CSST-2021-A02 and CMS-CSST-2021-A04. VAF acknowledges funding from an United Kingdom Research and Innovation grant (code:MR\/V022830\/1). Malgorzata Siudek acknowledges for the Polish National Agency for Academic Exchange (Bekker grant BPN/BEK/2021/1/00298/DEC/1), the European Union's Horizon 2020 Research and Innovation programme under the Maria Sklodowska-Curie grant agreement (No. 754510).

This material is based upon work supported by the U.S. Department of Energy (DOE), Office of Science, Office of High-Energy Physics, under Contract No. DE–AC02–05CH11231, and by the National Energy Research Scientific Computing Center, a DOE Office of Science User Facility under the same contract. Additional support for DESI was provided by the U.S. National Science Foundation (NSF), Division of Astronomical Sciences under Contract No. AST-0950945 to the NSF’s National Optical-Infrared Astronomy Research Laboratory; the Science and Technologies Facilities Council of the United Kingdom; the Gordon and Betty Moore Foundation; the Heising-Simons Foundation; the French Alternative Energies and Atomic Energy Commission (CEA); the National Council of Science and Technology of Mexico (CONACYT); the Ministry of Science and Innovation of Spain (MICINN), and by the DESI Member Institutions: \url{https://www.desi.lbl.gov/collaborating-institutions}.

The DESI Legacy Imaging Surveys consist of three individual and complementary projects: the Dark Energy Camera Legacy Survey (DECaLS), the Beijing-Arizona Sky Survey (BASS), and the Mayall z-band Legacy Survey (MzLS). DECaLS, BASS and MzLS together include data obtained, respectively, at the Blanco telescope, Cerro Tololo Inter-American Observatory, NSF’s NOIRLab; the Bok telescope, Steward Observatory, University of Arizona; and the Mayall telescope, Kitt Peak National Observatory, NOIRLab. NOIRLab is operated by the Association of Universities for Research in Astronomy (AURA) under a cooperative agreement with the National Science Foundation. Pipeline processing and analyses of the data were supported by NOIRLab and the Lawrence Berkeley National Laboratory. Legacy Surveys also uses data products from the Near-Earth Object Wide-field Infrared Survey Explorer (NEOWISE), a project of the Jet Propulsion Laboratory/California Institute of Technology, funded by the National Aeronautics and Space Administration. Legacy Surveys was supported by: the Director, Office of Science, Office of High Energy Physics of the U.S. Department of Energy; the National Energy Research Scientific Computing Center, a DOE Office of Science User Facility; the U.S. National Science Foundation, Division of Astronomical Sciences; the National Astronomical Observatories of China, the Chinese Academy of Sciences and the Chinese National Natural Science Foundation. LBNL is managed by the Regents of the University of California under contract to the U.S. Department of Energy. The complete acknowledgments can be found at \url{https://www.legacysurvey.org/}.

Any opinions, findings, and conclusions or recommendations expressed in this material are those of the author(s) and do not necessarily reflect the views of the U. S. National Science Foundation, the U. S. Department of Energy, or any of the listed funding agencies.

The authors are honored to be permitted to conduct scientific research on Iolkam Du’ag (Kitt Peak), a mountain with particular significance to the Tohono O’odham Nation.

We also acknowledge SDSS for providing extensive spectral database support. SDSS is managed by the Astrophysical Research Consortium for the Participating Institutions of the SDSS Collaboration including the Brazilian Participation Group, the Carnegie Institution for Science, Carnegie Mellon University, Center for Astrophysics | Harvard \& Smithsonian, the Chilean Participation Group, the French Participation Group, Instituto de Astrofísica de Canarias, The Johns Hopkins University, Kavli Institute for the Physics and Mathematics of the Universe(IPMU)/University of Tokyo, the Korean Participation Group, Lawrence Berkeley National Laboratory, Leibniz Institut fürAstrophysik Potsdam (AIP), Max-Planck-Institut für Astronomie (MPIA Heidelberg), Max-Planck-Institut für Astrophysik (MPA Garching), Max-Planck-Institut für ExtraterrestrischePhysik (MPE), National Astronomical Observatories of China, New Mexico State University, New York University, University of Notre Dame, Observatário Nacional/MCTI, The Ohio State University, Pennsylvania State University, Shanghai Astronomical Observatory, United Kingdom Participation Group, Universidad Nacional Autónoma de México, University of Arizona, University of Colorado Boulder, University of Oxford, University of Portsmouth, University of Utah, University of Virginia, University of Washington, University of Wisconsin, Vanderbilt University, and Yale University.

We acknowledge the efforts for public data from CTRS, PS1, PTF and ZTF. The Catalina Sky Survey  is funded by the National Aeronautics and Space Administration under Grant No. NNG05GF22G issued through the Science Mission Directorate Near-Earth Objects Observations Program. The CRTS survey is supported by the US National Science Foundation under grants AST-0909182 and AST-1313422. The CRTS survey is supported by the US National Science Foundation under grants AST-0909182 and AST-1313422. The PS1 has been made possible through contributions by the Institute for Astronomy, the University of Hawaii, the Pan-STARRS Project Office, the Max-Planck Society and its participating in- stitutes, the Max Planck Institute for Astronomy, Heidelberg and the Max Planck Institute for Extraterrestrial Physics, Garching, The Johns Hopkins University, Durham University, the University of Edinburgh, Queen’s University Belfast, the Harvard-Smithsonian Center for Astrophysics, the Las Cumbres Observatory Global Telescope Network Incorporated, the National Central University of Taiwan, the Space Telescope Science Institute, the National Aeronautics and Space Administration under Grant No. NNX08AR22G issued through the Planetary Science Division of the NASA Science Mission Directorate, the National Science Foundation under Grant No. AST-1238877, the University of Maryland, and Eotvos Lorand University (ELTE). PTF image data are  obtained with the Samuel Oschin Telescope and the 60 inch Telescope at the Palomar Observatory as part of the Palomar Transient Factory project, a scientific collaboration between the California Insti- tute of Technology, Columbia University, Las Cumbres Obser- vatory, the Lawrence Berkeley National Laboratory, the National Energy Research Scientific Computing Center, the University of Oxford, and the Weizmann Institute of Science. ZTF is supported by the National Science Foundation under Grant No. AST-2034437 and a collaboration including Caltech, IPAC, the Weizmann Institute for Science, the Oskar Klein Center at Stockholm University, the University of Maryland, Deutsches Elektronen-Synchrotron and Humboldt University, the TANGO Consortium of Taiwan, the University of Wis- consin at Milwaukee, Trinity College Dublin, Lawrence Livermore National Laboratories, and IN2P3, France. Operations are conducted by COO, IPAC, and UW.
\bibliographystyle{aasjournal}
\bibliography{ref}
\end{document}